\documentclass[12pt]{article}
\usepackage{amsmath,amsfonts,amssymb,amsthm,epsfig, graphicx, hyperref}
\usepackage[dvipsnames,table,dvipsnames*, svgnames*, hyperref]{xcolor}
\usepackage[linesnumbered,ruled]{algorithm2e}
\usepackage{natbib}
\usepackage{empheq}
\usepackage{caption}
\usepackage{subcaption}
\usepackage{babel,blindtext}
\usepackage{mwe}
\usepackage{float}
\usepackage{aliascnt}
\usepackage{mathrsfs,enumitem}
\usepackage{enumerate}
\usepackage{booktabs}
\usepackage{lscape}
\usepackage{setspace}

\DeclareMathSizes{12}{12}{5}{3}

\newcommand{\argmax}{\mathop{\rm \arg\!\max}}

\newcommand{\blind}{1}

\begin{document}

\sloppy

\def\spacingset#1{\renewcommand{\baselinestretch}%
{#1}\small\normalsize} \spacingset{1}

%%%%%%%%%%%%%%%%%%%%%%%%%%%%%%%%%%%%%%%%%%%%%%%%%%%%%%%%%%%%%%%%%%%%%%%%%%%%%%

\date{}

\if1\blind
{
  \title{\bf Sensitivity Analyses of Clinical Trial Designs: Selecting Scenarios and Summarizing Operating Characteristics
  
  % Optimal Selection of Scenarios to Illustrate Variations in Operating Characteristics
  
  % Representative and Optimal Sensitivity Analysis (ROSA) of Clinical Trial Designs
  }
\author{Larry Han$^{1}$, Andrea Arf\`e$^{2}$, Lorenzo Trippa$^{1,3}$ \\ \\
    1 Department of Biostatistics, \\ Harvard T.H. Chan School of Public Health \\ \\
    2 Department of Epidemiology and Biostatistics, \\ Memorial Sloan Kettering Cancer Center \\ \\
    3 Department of Biostatistics and Computational Biology, \\ Dana-Farber Cancer Institute
}
  \maketitle
} \fi

\if0\blind
{
  \bigskip
  \bigskip
  \bigskip       
  \begin{center}
    {\LARGE \bf  ROSA}
\end{center}
  \medskip
} \fi

\bigskip

\begin{abstract}

The use of simulation-based sensitivity analyses is fundamental to evaluate and compare candidate designs for future clinical trials. In this context, sensitivity analyses are especially useful to assess the dependence of important design operating characteristics (OCs) with respect to various unknown parameters (UPs). % such the enrollment rate or the magnitude of potential treatment effects.   
Typical examples of OCs include the likelihood of detecting treatment effects  and the average study duration, which depend on UPs that are not known until after the onset of the clinical study, such as the distributions of the primary outcomes and patient profiles. Two crucial components of sensitivity analyses are (i) the choice of a set of plausible simulation scenarios $\{\boldsymbol{\theta}_1,...,\boldsymbol{\theta}_K\}$ and (ii) the list of OCs of interest. We propose a new approach to choose the set of scenarios for inclusion in design sensitivity analyses. Our approach balances the need for simplicity and interpretability of OCs computed across several scenarios with the need to faithfully summarize---through simulations---how the OCs vary across all plausible values of the UPs. Our proposal also supports the selection of the number of simulation scenarios to be included in the final sensitivity analysis report. To achieve these goals, we minimize a loss function $\mathcal{L}(\boldsymbol{\theta}_1,...,\boldsymbol{\theta}_K)$ that formalizes whether a specific set of $K$ sensitivity  scenarios $\{\boldsymbol{\theta}_1,...,\boldsymbol{\theta}_K\}$ is adequate to summarize how the OCs of the trial design vary across all plausible values of the UPs. Then, we use optimization techniques to select the best set of simulation scenarios to exemplify the OCs of the trial design.  

\end{abstract}

\noindent%
{\it Keywords: Clinical trial design; operating characteristics; sensitivity analysis; function approximation; simulated annealing.}  
\vfill

\spacingset{1.9} % DON'T change the spacing!

\section{Introduction}
\label{s:intro}

% Modern designs have multiple objectives
Clinical trial designs are becoming increasingly complex to meet the multifaceted needs and goals of precision medicine. 
Examples of complex designs include adaptive seamless phase i/ii designs for evaluating, early in the treatment development process, the dosing, safety, and activity of new drugs \citep{hobbs2019seamless}. Also, adaptive randomized trials with frequent interim looks at the data can evaluate one or more therapies simultaneously while attempting to minimize trial duration and resources \citep{thorlund2018key, berry2010bayesian}. Additional examples of complex designs have been implemented in biomarker-stratified trials to evaluate the efficacy of a therapy and possible variations of treatment effects across patient subgroups \citep{mehta2019adaptive}. 

%We care about the multiple OCs of a trial design
When planning a new trial, it is necessary to predict and evaluate several operating characteristics (OCs) $\mathbf{f}=(f_1,\ldots,f_R)$. Relevant OCs can include the likelihood of selecting an effective dose with low toxicity in a  phase i/ii study, the probability of detecting treatment effects in a randomized study, the expected trial duration, costs, and other metrics to evaluate designs that often enroll patients from  different subgroups. Multiple OCs typically need to be examined jointly in order to evaluate the relevant trade-offs achieved by candidate designs, such as balancing  the accuracy in estimating treatment effects and the expected study duration.

%OCs are determined by uncertain parameters
The obvious challenge for evaluating a candidate design is that the value of $\mathbf{f}$, the vector of OCs of the study design, is not known, and it is difficult to estimate before the onset of the  trial. Indeed, $\mathbf{f}$ is usually a function $\mathbf{f}=\mathbf{f}(\boldsymbol{\theta})$ of a vector $\boldsymbol{\theta}=(\theta_1, \ldots, \theta_d)$ that contains unknown parameters (UPs) which identify the distribution of all relevant variables that will be captured during the trial. For example, UPs can include the enrollment and drop-out rates, the magnitude of treatment effects, and the prevalence of  predictive biomarkers in the trial population. Uncertainty on $\boldsymbol{\theta}$ makes it non-trivial to evaluate whether a candidate design is appropriate for implementing the new study.

%Standard approach for sensitivity analyses of OCs 
Sensitivity analyses are commonly used to account for uncertainty on UPs $\boldsymbol{\theta}$ and OCs $\mathbf{f}(\boldsymbol{\theta})$ when evaluating a candidate design.
 They typically proceed in three steps.
  First, a set of plausible scenarios $\{\boldsymbol{\theta}_1,\ldots,\boldsymbol{\theta}_K\}$, i.e., specific values of the vector $\boldsymbol{\theta}$ of UPs, is selected. 
Next, the corresponding set of OCs are computed $\{\boldsymbol{\mathbf{f}}(\boldsymbol{\theta}_1)$, $\ldots$, $\boldsymbol{\mathbf{f}}(\boldsymbol{\theta}_K)\}$ using trial simulations or analytic results.
 Finally, based on the computed OCs and their variations across the set of $K$ scenarios, the investigators evaluate if the candidate design is appropriate to achieve the  aims of the study.
 Throughout the manuscript, we use the terms {\it sensitivity analysis} or {\it sensitivity report} to indicate a set of $K$ scenarios $\{\boldsymbol{\theta}_1,...,\boldsymbol{\theta}_K\}$ and the associated OCs $\{\boldsymbol{\mathbf{f}}(\boldsymbol{\theta}_1)$, $\ldots$, $\boldsymbol{\mathbf{f}}(\boldsymbol{\theta}_K)\}$ which are computed to illustrate how the OCs $\mathbf{f}(\boldsymbol{\theta})$ vary across plausible values of UPs $\boldsymbol{\theta}$.

%Problems with sensitivity analyses
Producing a sensitivity report to effectively evaluate a study design presents several challenges.
Indeed, it can be difficult to select the set of UPs $\{\boldsymbol{\theta}_1,\ldots,\boldsymbol{\theta}_K\}$, especially if the dimension $d$ of the vector $\boldsymbol{\theta}$ is moderate to high (say $d\ge 5$). For the investigators, it might be unclear if the selected scenarios are adequate to illustrate the variations  of the OCs $\mathbf{f}$ across potential values  of the UPs $\boldsymbol{\theta}$. Similarly, for regulators, there may be skepticism as to whether the selected scenarios are chosen to highlight positive aspects of the trial design without pointing at its limitations and negative aspects  \citep{razavi2021future}. Another  subtle challenge is the choice of the number of scenarios $K$.
%how many scenarios to choose. 
Indeed, a large number of scenarios (say $K=100$)  may simplify the task of representing how the OCs $\mathbf{f}(\boldsymbol{\theta})$ vary across potential values of the UPs $\boldsymbol{\theta}$, but a sensitivity report that contains too many scenarios
makes it difficult to interpret and communicate the included results.

A common approach used to define the set of scenarios $\{\boldsymbol{\theta}_1,...,\boldsymbol{\theta}_K\}$ is to vary a single entry of $\boldsymbol{\theta}$ (say $\theta_1$) while fixing the other UPs to some reference values $\hat{\theta}_2,...,\hat{\theta}_d$, which might be estimated from previous studies. In this case, the set of scenarios becomes $\{ \boldsymbol{\theta}_1 =({\theta_{1,1},\hat{\theta}_2,...,\hat{\theta}_d})),...,{\boldsymbol{\theta}_K}= ({\theta_{K,1},\hat{\theta}_2,...,\hat{\theta}_d})\}$. Similar perturbations of $\boldsymbol{\hat{\theta}}$ can be repeated for the other entries of the vector. 
However, this approach can be inadequate if the relation between
 one UP ($\theta_1$, say) and the OCs changes when we consider distinct values of the other UPs ($\theta_2,...,\theta_d$). In such cases, this approach compromises the possibility of illustrating with sufficient accuracy how the OCs $\mathbf{f}(\boldsymbol{\theta})$ vary across plausible values of the UPs $\boldsymbol{\theta}$, for example  
 through a table of OCs  $\mathbf{f}(\boldsymbol{\theta}_1),...,\mathbf{f}(\boldsymbol{\theta}_K)$ computed for a
 representative set of scenarios $\{\boldsymbol{\theta}_1,...,\boldsymbol{\theta}_K\}$.

%Our appoach to sensitivity analysis
We propose a method to choose an optimal set of scenarios $\{\boldsymbol{\theta}_1^*,...,\boldsymbol{\theta}_K^*\}$  for a sensitivity  report
that will provide relevant  OCs $\{\mathbf{f}(\boldsymbol{\theta}_1^*),...,\mathbf{f}(\boldsymbol{\theta}_K^*)\}$.
This decision is based on a formal utility criterion $\mathcal{U}$. 
The function $\mathcal{U} : \{\boldsymbol{\theta}_1,...,\boldsymbol{\theta}_K\} \mapsto \mathbb{R}$ formalizes the ability of any set of scenarios $\{\boldsymbol{\theta}_1,...,\boldsymbol{\theta}_K\}$ to represent the  map $\boldsymbol{\Theta} \rightarrow {\mathbf{f}}(\boldsymbol{\Theta})$, where $\boldsymbol{\Theta}$
is the parameter space. In some cases, we will consider a   restriction $\boldsymbol{\Theta}^\prime \subset \boldsymbol{\Theta}$  to focus only on plausible  values of the UPs $\boldsymbol{\theta}$.  
This utility criterion $\mathcal{U}$ assigns high (low) utility to  $\{\boldsymbol{\theta}_1,...,\boldsymbol{\theta}_K\}$ if  
   the table of potential UPs and OCs
$ [(\boldsymbol{\theta}_1,\boldsymbol{\mathbf{f}}(\boldsymbol{\theta}_1))$, $\ldots$, $(\boldsymbol{\theta}_K,\boldsymbol{\mathbf{f}}(\boldsymbol{\theta}_K))]$ can be viewed as an accurate  (inaccurate) summary of how the design's OCs vary across  $\boldsymbol\Theta$ or the restricted parameter space $\boldsymbol{\Theta}^\prime$. We call the set of scenarios that maximizes $\mathcal{U}$ the Representative and Optimal Sensitivity Analysis (ROSA) scenarios.
  To select ROSA scenarios $\{\boldsymbol{\theta}_1^*, \ldots, \boldsymbol{\theta}_K^*\}$, we introduce a computational procedure that leverages (i) flexible regression  methods like neural networks (NNs) \citep{goodfellow2016deep} 
  %or tree-based models \citep{chipman2010bart} 
  and (ii) optimization algorithms like simulated annealing (SA) \citep{belisle1992convergence}. Our approach is applicable to any trial design, regardless of the number $d$ of UPs in $\boldsymbol{\theta}$ and the number $R$ of OCs in $\mathbf{f}(\boldsymbol{\theta})$. 

%Structure of the paper
To illustrate the method, we conduct sensitivity analyses for three trial designs. 
The first is a two-arm randomized design that aims to test and estimate the effects of an experimental treatment compared to the standard of care (SOC).
The second is a multi-stage randomized trial that leverages an auxiliary outcome $S$ measured shortly after randomization for interim decisions and a primary outcome $Y$ with a longer ascertainment time \citep{niewczas2019interim}.
The third is a biomarker-adaptive enrichment design similar to the design of the TAPPAS trial \citep{mehta2019adaptive}, a randomized phase iii  trial comparing TRC105 and pazopanib versus pazopanib alone in patients with advanced angiosarcoma \citep{jenkins2011adaptive, jones2017tappas}. In the first design, we consider a single UP and a single OC, whereas the latter two designs consider multiple UPs and multiple OCs.

\section{Selecting sensitivity  scenarios}
We  introduce our procedure to select $K$ sensitivity scenarios $\boldsymbol{\theta}_1,...,\boldsymbol{\theta}_K \in\boldsymbol{\Theta}$, where $\boldsymbol{\Theta}$ is the set of  potential values 
 of the UPs  $\boldsymbol{\theta}$.  
We assume that $\boldsymbol{\Theta}$ is a  bounded subset of $\mathbb{R}^d$
 and use the notation $||\cdot||_2$ to indicate the Euclidean norm on $\mathbb{R}^d$.  We will restrict $\boldsymbol{\Theta}$ to a subset  $\boldsymbol{\Theta}^\prime$  when there is sufficient  prior information from completed studies or clinical experience.  
 
\subsection{Optimization of a utility criterion}\label{seq:optfrm}
We identify ROSA scenarios $\boldsymbol{\theta}_1^*,...,\boldsymbol{\theta}_K^*$ as the scenarios 
  that maximize a utility criterion $\mathcal{U}$
\begin{equation}
    \boldsymbol{\theta}_1^*,...,\boldsymbol{\theta}_K^* = \argmax_{\boldsymbol{\theta}_1,...,\boldsymbol{\theta}_K} \text{ } \mathcal{U}(\boldsymbol{\theta}_1,...,\boldsymbol{\theta}_K),
\end{equation}
where
\begin{equation} \label{eq:loss}
\mathcal{U}(\boldsymbol{\theta}_1,...,\boldsymbol{\theta}_K) = -\max_{\boldsymbol{\theta}^\prime \in \boldsymbol{\Theta}} \left\{\min_{k = 1,...,K} D[\mathbf{f}(\boldsymbol{\theta}^\prime),\mathbf{f}(\boldsymbol{\theta}_k)] \right\}.
\end{equation} 
We can  symmetrically define  the corresponding loss function 
  $\mathcal{L}= -\mathcal{U}$ by inverting the sign in equation (\ref{eq:loss}).
Here, $D[\mathbf{f}(\boldsymbol{\theta}^\prime),\mathbf{f}(\boldsymbol{\theta}_k)]$ is a metric between the OCs   $\mathbf{f}(\boldsymbol{\theta}^\prime)=(f_1(\boldsymbol{\theta}^\prime), \ldots f_R(\boldsymbol{\theta}^\prime))$ and $\mathbf{f}(\boldsymbol{\theta}_k)=(f_1(\boldsymbol{\theta}_k,)\ldots,f_R(\boldsymbol{\theta}_k))$. We will consider metrics of the form
$$D[{\textbf{f}}(\boldsymbol{\theta}^\prime),{\textbf{f}}\left(\boldsymbol{\theta_k})\right] =  \sum_{r=1}^R w_r ||f_r(\boldsymbol{\theta}^\prime)-f_r(\boldsymbol{\theta}_k)||_2,$$
where $w_1, \ldots, w_R$ are  non-negative weights that sum to one.
% We use these weights to express all OCs on a common measurement scale, as well as to calibrate their relative importance. 

% Interpretation: we search for the set of scenarios that provides the lowest approximation error
We can now provide an explicit interpretation of the utility function $\mathcal{U}$ in equation (\ref{eq:loss}) and  of the  loss function $\mathcal{L}=-\mathcal{U}$.
Consider a set of scenarios $\{\boldsymbol{\theta}_1,...,\boldsymbol{\theta}_K\}$ -- the order of the entries
 is not relevant -- and $\boldsymbol{\theta}^\prime$ in $\boldsymbol{\Theta}$. 
 For  $1 \leq k \leq K$, the metric $D[\mathbf{f}(\boldsymbol{\theta}^\prime), \mathbf{f}(\boldsymbol{\theta}_k)]$ is a summary of the  differences  between the  OCs $\mathbf{f}$ at  $\boldsymbol{\theta}^\prime$ and  
  the same OCs  when we consider  the $k$-th scenario $\boldsymbol{\theta}_k$. 
 Therefore, $\min_{k = 1,...,K} D[\mathbf{f}(\boldsymbol{\theta}^\prime),\mathbf{f}(\boldsymbol{\theta}_k)]$ 
 can be viewed as an approximation error   between
 $\mathbf{f}(\boldsymbol{\theta}^\prime)$ 
 and  a similar vector of OCs selected among our $K$ options
 $\mathbf{f}(\boldsymbol{\theta}_1),...,\mathbf{f}(\boldsymbol{\theta}_K)$. 
 Expression (\ref{eq:loss}) identifies through the maximization operator the
  worst-case, with highest approximation error, 
that we can obtain  by varying $\boldsymbol{\theta}^\prime$  in $\boldsymbol{\Theta}$. 
%is equal to $\mathcal{L}(\boldsymbol{\theta}_1,...,\boldsymbol{\theta}_K)$, the value of the loss function.
 We  maximize the utility function $\mathcal{U} $ and  use $\boldsymbol{\theta}_1^*,...,\boldsymbol{\theta}_K^*$ to 
  indicate the ROSA scenarios. Alternative utility criteria and loss functions are described later in the manuscript.

\begin{figure}[H]
    \centering
    \includegraphics[scale=0.45]{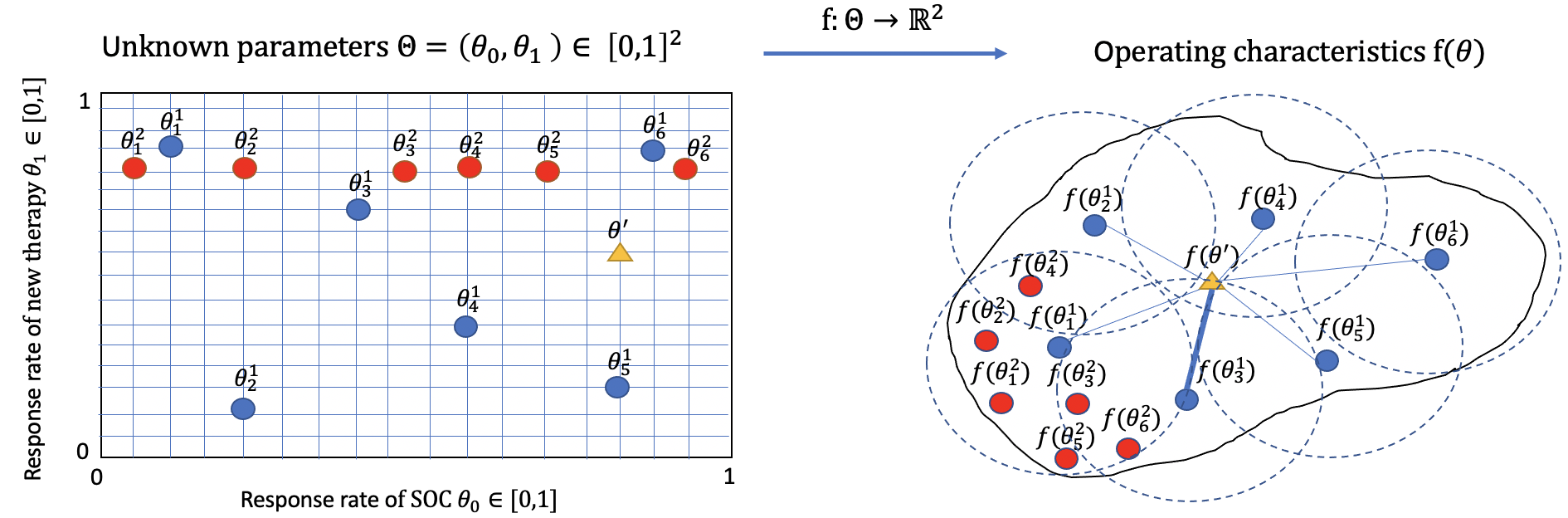}
      \caption{Geometric representation of an arbitrary scenario $\boldsymbol{\theta}^\prime$ and two proposed sets of scenarios. (Left) Parameter space $\boldsymbol{\Theta} =  [0,1]^2$ with arbitrary scenario $\boldsymbol{\theta}^\prime$ {\color{orange}{(orange triangle)}} and two sets of proposed scenarios $\{\boldsymbol{\theta}_1^1,...,\boldsymbol{\theta}_6^1\}$ {\color{blue}{(blue points)}} and  $\{\boldsymbol{\theta}_1^2,...,\boldsymbol{\theta}_6^2\}$ {\color{red}{(red points)}}. (Right) OC surface $\mathbf{f}(\boldsymbol{\Theta})$ with the corresponding  OCs for $\boldsymbol{\theta}^\prime$ and the two proposed sets of scenarios. The radius of the dotted circles (with blue points as centers) is the value of the loss $\mathcal{L}$ associated with the blue points. ROSA scenarios minimize the loss $\mathcal{L}$, which in turn is equal to the radius of the dotted circles that cover the OC surface $\mathbf{f}(\boldsymbol{\Theta})$. }
    \label{fig:fig1}
\end{figure}

To provide a geometric interpretation of the loss function $\mathcal{L}$, we illustrate how one set of $K$ scenarios can be preferable to a different set of $K$ scenarios (Figure \ref{fig:fig1}). Specifically, suppose we aim to design a single-arm trial with an interim analysis that allows for early-stopping for futility. The goal of the trial is to compare the response rate of an experimental drug $\theta_1$ with that of the SOC $\theta_0$ at the end of the study. However, because study patients only receive the experimental drug, the response rate under the SOC $\theta_0$ is estimated ($\widehat{\theta}_0$) before the onset of the study, for example using data from a previous trial. At the interim analysis, the trial may stop for futility if the preliminary evidence of positive treatment effects $\Delta_{interim}$ is insufficient to continue the study.
During the final analysis, the null hypothesis $H_0: \theta_1 \le \widehat{\theta}_0$ (the experimental therapy is not superior to the historical control) is tested against the alternative hypothesis $H_1: \theta_1 > \widehat{\theta}_0$ (the experimental therapy is superior to the historical control). The discrepancy between $\theta_0$ and $\widehat{\theta}_0$ could impact the OCs, such as the probability of  stopping the study for futility \citep{vanderbeek2019randomize}. In this design, $\boldsymbol{\theta} = (\theta_0,\theta_1)$ are the UPs, and $\boldsymbol{\Theta}=[0,1]^2$. 
Suppose that there are two OCs of interest:
(i) $f_1$, power, defined as the probability of a true positive result when the experimental
drug has beneficial effects compared to the SOC ($f_1$ is equal to zero if the treatment effects are null or negative) and (ii) $f_2$, the expected sample size. 

The left panel of Figure \ref{fig:fig1} is a  representation of $\boldsymbol{\Theta}$. We are interested in the two OCs of the single-arm design. Two sets of $K = 6$ scenarios are proposed. The first set of scenarios $\{\boldsymbol{\theta}_1^1,...,\boldsymbol{\theta}_6^1\}$ (blue points) is chosen by varying both UPs at the same time, while the second set $\{\boldsymbol{\theta}_1^2,..., \boldsymbol{\theta}_6^2\}$ (red points) is chosen by varying only $\theta_0$ while fixing the value of $\theta_1$. The two sets of scenarios, the corresponding OCs, and associated loss $\mathcal{L}$ are represented  in the right panel of Figure \ref{fig:fig1}. The first set of scenarios (blue points) is preferred over the second set of scenarios (red points) because it is more representative of the variation of the OCs over $\boldsymbol{\Theta}$. Geometrically, the loss $\mathcal{L}(\boldsymbol{\theta}_1^1,...,\boldsymbol{\theta}_6^1)$ associated with the blue points is identical to the  minimum radius of the circles with centers $\mathbf{f}(\boldsymbol{\theta}_1^1),...,\mathbf{f}(\boldsymbol{\theta}_6^1)$  (see Figure \ref{fig:fig1})
 necessary
 to cover the OCs surface $\mathbf{f}(\boldsymbol{\Theta})$. 

\subsection{Estimating the OCs}
We  describe an algorithm to numerically approximate the OCs $\mathbf{f}(\boldsymbol{\theta})$ for every $\boldsymbol{\theta}\in \boldsymbol{\Theta}$. 
This is necessary to solve the optimization problem in equation (\ref{eq:loss}). 
Indeed, in most cases the function $\mathbf{f}(\boldsymbol{\theta})$ cannot be computed in closed form. 

We briefly outline our four-step procedure. In the first step,
 we choose a  large number $J$ (say $J = 1000$) of  training scenarios 
 $\boldsymbol{\theta}_1^t,...,\boldsymbol{\theta}_J^t.$
In the second step,
   we use Monte Carlo simulations to obtain  estimates 
   $\bar{\mathbf{f}}(\boldsymbol{\theta}_1^t)$, ..., $\bar{\mathbf{f}}(\boldsymbol{\theta}_J^t)$ of 
   $\mathbf{f}(\boldsymbol{\theta}_1^t)$, ..., $\mathbf{f}(\boldsymbol{\theta}_J^t)$. 
   In the third step, we train a flexible regression model -- we use NNs in our implementation -- based on the data points 
   $(\boldsymbol{\theta}_1^t, \bar{\mathbf{f}}(\boldsymbol{\theta}_1^t))$, ..., $(\boldsymbol{\theta}_J^t, \bar{\mathbf{f}}(\boldsymbol{\theta}_J^t))$.  The output of this step is a regression function $\hat{\mathbf{f}}(\boldsymbol{\theta})$ that is easy to compute at any $\boldsymbol{\theta}\in \boldsymbol{\Theta}$ and that approximates  $\mathbf{f}(\boldsymbol{\theta})$. 
   In the fourth step, we validate the  regression model based on $J^\prime$ (say $J^\prime = 200$)
   independent  simulations  
   $(\boldsymbol{\theta}_1^v, \bar{\mathbf{f}}(\boldsymbol{\theta}_1^v))$, ..., $(\boldsymbol{\theta}_{J^\prime}^v, \bar{\mathbf{f}}(\boldsymbol{\theta}_{J^\prime}^v))$. 
 Steps 1-3 of this procedure are summarized in Algorithm 1. Step 4 is described in Algorithm 2. 

In more detail, in step 1,
to select the training scenarios $\boldsymbol{\theta}_1^t,...,\boldsymbol{\theta}_J^t$,
 we randomly select $J$ scenarios in $\boldsymbol{\Theta}$ 
 using Latin hypercube sampling (LHS) \citep{mckay2000comparison, carnell2016package}. 
 LHS generates $J$ scenarios by first partitioning the $d$ UP dimensions into $J$ non-overlapping intervals and selecting one value from each interval at random. The $J$ values obtained for the first UP ${\theta}_1$ 
 are randomly paired with the $J$ values obtained for the second ${\theta}_2$, and so on, for all $d$ UPs to form $J$ $d$-tuples, which constitute the training scenarios $\boldsymbol{\theta}_1^t,...,\boldsymbol{\theta}_J^t$.

In step 2, we estimate the OCs of the trial design. For simplicity, we  consider OCs defined as expected values (e.g., bias, power, duration of the trial, etc.),  but the algorithm  can  be easily modified to consider other OCs. Specifically, we assume that ${\mathbf{f}}(\boldsymbol{\theta}) = \mathbb{E}_{\boldsymbol{\theta}}[\boldsymbol{\varphi}(Z, \boldsymbol{\theta})]$ for some function $\boldsymbol{\varphi}$, where the random vector $Z$ represents the data generated during the trial -- including the collection of treatment assignment indicators and realized patient outcomes -- under scenario $\boldsymbol{\theta}$. For example, $\boldsymbol{\varphi}$ can be the indicator that captures if a null hypothesis of interest has been correctly rejected at the end of the study, or the duration of the simulated trial. In practice, to estimate ${\mathbf{f}}(\boldsymbol{\theta})$, we proceed as follows. First, for each of the training scenarios $\boldsymbol{\theta}_j^t$, $1 \leq j \leq J$, we simulate
$M$ (say $M = 200$) clinical trials following the trial design.  We then use the $M$ scenario-specific  simulated trials to compute the estimate 
$$\bar{\mathbf{f}}(\boldsymbol{\theta}_j^t) = M^{-1} \sum_{m=1}^M \boldsymbol{\varphi}(Z_{j,m}, \boldsymbol{\theta}_j^t), \quad 1 \leq j \leq J,$$ where $Z_{j,m}$ is the $m^{th}$ trial dataset simulated under the $j^{th}$ training scenario $\boldsymbol{\theta}_j^t$. 

In step 3, we have only two inputs, the scenarios 
$\boldsymbol{\theta}_j^t$ and the  estimates $\bar{\mathbf{f}}(\boldsymbol{\theta}_j^t)$, $1 \leq j \leq J$, to fit a function  $\hat{\mathbf{f}}(\boldsymbol{\theta})$. For example, one could  use NNs, splines \citep{bookstein1989principal}, or Gaussian processes \citep{rasmussen2003gaussian}.
We use NN regression functions in our applications because these are easy to compute using widely available software and have been demonstrated to have good empirical performance \citep{leshno1993multilayer, hornik1991approximation, goodfellow2016deep}. 
 %For simplicity, we only considered NNs with $3$ hidden layers ($8$, $64$, and $64$ neurons, respectively) with ReLU activation functions. We fit all NNs using the Keras implementation of the Adam optimization algorithm \citep{arnold2017kerasr}.]

\vspace{\baselineskip}
 \begin{algorithm}[H]
 % \KwData{Parameter space $\boldsymbol{\Theta}$, Trial design}
 \hspace*{0em} \textbf{Input:} Trial design, Parameter space $\boldsymbol{\Theta}$, $J$, $M$

 \underline{Step 1}: Select $J$ scenarios $\boldsymbol{\theta}_1^t,...,\boldsymbol{\theta}_J^t \in \boldsymbol{\Theta}$ \\

 \underline{Step 2}: \For{$j=1$ to $J$}{
    Simulate $M $ trials  \\
    Obtain approximate OCs  $\boldsymbol{\bar{f}}(\boldsymbol{\theta}_j^t) = M^{-1} \sum_{m=1}^M \boldsymbol{\varphi}(Z_{j,m}, \boldsymbol{\theta}_j^t)$,
     where $\boldsymbol{\varphi}$ is a function of $Z_{j,m}$, the $m^{th}$ trial dataset simulated under the $j^{th}$ scenario, and the corresponding parameter $\boldsymbol{\theta}_j$
   }

 \underline{Step 3}: Obtain an approximation of the OCs $\hat{\boldsymbol{f}}$ by training a regression algorithm, for example a NN model, and  using  the data points $(\boldsymbol{\theta}_j^t, \boldsymbol{\bar{f}}(\boldsymbol{\theta}_j^t))$, $1 \leq j \leq J$  \\

   \hspace*{0em} \textbf{Output:} Function $\hat{\mathbf{f}}(\boldsymbol{\theta})$ 
 \caption{ Obtaining a function $\boldsymbol{\hat{f}}$ that approximates the OC function $\mathbf{f}$}
\end{algorithm} 
\vspace{\baselineskip}

  In step 4 (Algorithm 2), we investigate the differences between  $\hat{\boldsymbol{f}}$ and $\boldsymbol{f}$.
 Specifically, we first select at random  $J^\prime$  validation scenarios $\boldsymbol{\theta}_1^v,...,\boldsymbol{\theta}_{J^\prime}^v$
 independently with respect to previous computations (step 1-3) and simulate  $M^\prime$ trials  (say $M^\prime = 500$) for each $\boldsymbol{\theta}_{j^\prime}^v$, $1 \leq j^\prime \leq J^\prime$. 
 Based on the results of the simulated trials, for each $j^\prime$, we then  compute Monte Carlo  estimates $\bar{\mathbf{f}}(\boldsymbol{\theta}_{j^\prime}^v) = M^{\prime -1} \sum_{m^\prime = 1}^{M\prime}
 \boldsymbol{\varphi}(Z_{j^\prime,m^\prime}, \boldsymbol{\theta}_{j^\prime}^v)$ 
 of the OCs $\mathbf{f}(\boldsymbol{\theta}_{j^\prime}^v)$. For several important OCs (e.g., average sample size, expected duration, power, type 1 error), the estimator $\bar{\mathbf{f}}(\boldsymbol{\theta}_{j^\prime}^v) = M^{\prime -1} \sum_{m^\prime = 1}^{M\prime}
 \boldsymbol{\varphi}(Z_{j^\prime,m^\prime}, \boldsymbol{\theta}_{j^\prime}^v)$ is unbiased.
 Finally, we compare the estimates
  $\bar{\boldsymbol{f}}(\boldsymbol{\theta}_{j^\prime}^v)$ 
  and  the independent  estimates $\hat{\boldsymbol{f}}(\boldsymbol{\theta}_{j^\prime}^v)$.
   We use summary statistics and graphs to evaluate the 
   differences 
    $\hat{\boldsymbol{f}}(\boldsymbol{\theta}_{j^\prime}^v) - \bar{\boldsymbol{f}}(\boldsymbol{\theta}_{j^\prime}^v)$.
If the approximation  $\hat{\boldsymbol{f}}(\boldsymbol{\theta}_{j^\prime}^v)$  is not adequate, we can use a different regression methodology, increase the number $(M, M^\prime)$ of trials, or increase the number $J$ of training scenarios in Algorithm 1.

 \vspace{\baselineskip}
 \begin{algorithm}[H]
 \hspace*{0em} \textbf{Input:} 
 Approximation of the OCs $\hat{\boldsymbol{f}}$, Trial design, $J^\prime$
 
 Randomly select $J^\prime $ scenarios 
 $\boldsymbol{\theta}_1^v,...,\boldsymbol{\theta}_{J^\prime}^v \in \boldsymbol{\Theta}$ independently  
  from previous computations (Algorithm 1)\\
 
 \For{$j^\prime=1$ to $J^\prime$}{
 Simulate $M^\prime$  trials $Z_{j^\prime,m^\prime}$  \\
    Compute 
    $\bar{\boldsymbol{f}}(\boldsymbol{\theta_{j^\prime}^v})  = M^{\prime -1} \sum_{m^\prime=1}^{M^\prime} \boldsymbol{\varphi}( Z_{j^\prime,m^\prime}, \boldsymbol{\theta}_{j^\prime}^v)$ \\
     Compute $\hat{\boldsymbol{f}}(\boldsymbol{\theta}_{j^\prime}^v)$ } 
    \hspace*{0em} \textbf{Output:} Set of differences  $\hat{\boldsymbol{f}}(\boldsymbol{\theta}_{j^\prime}^v) - \bar{\boldsymbol{f}}(\boldsymbol{\theta}_{j^\prime}^v)$ and  scatterplots  to jointly visualize  the OC estimates $\bar{\boldsymbol{f}}(\boldsymbol{\theta}_{j^\prime}^v)$  and the independent estimates  $\hat{\boldsymbol{f}}(\boldsymbol{\theta}_{j^\prime}^v)$, $1 \leq j^\prime \leq J^\prime$.
    Compute summaries of the differences (e.g., median, range, or other descriptive statistics).
   % \vspace{0.05cm}

 {\bf Interpretation:} Differences between 
   $\bar{\boldsymbol{f}}(\boldsymbol{\theta}_{j^\prime}^v)$  and the independent estimates  $\hat{\boldsymbol{f}}(\boldsymbol{\theta}_{j^\prime}^v)$, $1 \leq j^\prime \leq J^\prime$, consistently close to zero  provide evidence that $\hat{\boldsymbol{f}}$  is an accurate approximation of $\boldsymbol{f}$
    \caption{Validating the approximation of the OCs $\hat{\boldsymbol{f}}$}
   \end{algorithm} 
\vspace{\baselineskip}

\subsection{Approximating the Loss Function}
After computing $\hat{\boldsymbol{f}}$ (Algorithm 1) and validating its accuracy (Algorithm 2), we use it to approximate the loss function  $\mathcal{L}(\boldsymbol{\theta}_1,...,\boldsymbol{\theta}_K)$. To proceed, we choose a diffuse and finite subset of the parameter space $\boldsymbol{\Theta}^F \subset \boldsymbol{\Theta}$. 
For example $\boldsymbol{\Theta}^F$ can include 100,000 random points from a distribution with support $\boldsymbol{\Theta}$. When $\boldsymbol{\Theta}^F$ contains a large number of random points that are distributed over  $\boldsymbol{\Theta}$, under minimal assumptions (e.g., compact $\boldsymbol{\Theta}$ and OCs with bounded range),
\begin{align*}
    \mathcal{L}(\boldsymbol{\theta}_1,...,\boldsymbol{\theta}_K) &=  \max_{\boldsymbol{\theta} \in \boldsymbol{\Theta}} \left\{\min_{k=1,...,K} D[\boldsymbol{f}(\boldsymbol{\theta}),\boldsymbol{f}(\boldsymbol{\theta}_k) ]\right\} \\
    &\approx \max_{\boldsymbol{\theta}^\prime \in \boldsymbol{\Theta}^F} \left\{\min_{k=1,...,K} D[\boldsymbol{\hat{f}}(\boldsymbol{\theta}^\prime),\boldsymbol{\hat{f}}(\boldsymbol{\theta}_k) ]\right\} = \hat{\mathcal{L}}(\boldsymbol{\theta}_1,...,\boldsymbol{\theta}_K).
\end{align*}
To summarize, we can approximate the loss function $\mathcal{L}(\boldsymbol{\theta}_1,...,\boldsymbol{\theta}_K)$ over the entire parameter space $\boldsymbol{\Theta}$ by $\hat{\mathcal{L}}(\boldsymbol{\theta}_1,...,\boldsymbol{\theta}_K)$ using a diffuse and finite subset $\boldsymbol{\Theta}^F$.

\subsection{Optimization}
We now aim to approximately minimize the loss function $\hat{\mathcal{L}}$.
To illustrate the need for approximate solutions, consider the setting of a single UP $(d = 1)$, a finite ${\boldsymbol{\Theta}}$, and an easy-to-compute loss function $\mathcal{L}$.  Even in this simple setting, identifying  $\boldsymbol{\theta}_1^*,...,\boldsymbol{\theta}_K^* \in \boldsymbol{\Theta}$ can be challenging. 
 For example, to select $K=10$ representative scenarios $\boldsymbol{\theta}_1^*,...,\boldsymbol{\theta}_K^*$ from $1000$ points $\{\boldsymbol{\theta}_j; 1 \leq j \leq 1000\}= \boldsymbol{\Theta}$,  the loss function $\hat{\mathcal{L}}$ would need to  be calculated for $2.63 \times 10^{23}$ different possible sets $\{\boldsymbol{\theta}_1,...,\boldsymbol{\theta}_K\}$.  In what follows, we describe the use of SA (Algorithm 3), a simple  strategy to  reduce the outlined computational burden, regardless if  $\boldsymbol{\Theta}$
is finite or not \cite{kirkpatrick1983optimization, belisle1992convergence, spall2005introduction}.

The SA algorithm proceeds as follows. First, initial scenarios $\boldsymbol{\theta}_{1}^1,...,\boldsymbol{\theta}_{K}^1$ are proposed, for example by sampling $\boldsymbol{\theta}_{1}^1,...,\boldsymbol{\theta}_{K}^1$
 from a probability distribution with support $\boldsymbol{\Theta}$. 
Then, iteratively for $1 \leq i \leq I$, the current scenarios $\boldsymbol{\theta}_{1}^{i},...,\boldsymbol{\theta}_{K}^{i}$ are perturbed by adding to them Gaussian noise variables $\boldsymbol{z}_{1}^i,...,\boldsymbol{z}_{K}^i$, thus obtaining new proposed scenarios $\boldsymbol{\theta}_{1}^\prime,...,\boldsymbol{\theta}_{K}^\prime$ (this step is represented by the ``Perturb'' operator in Algorithm 3).
 At each iteration, the proposed scenarios $\boldsymbol{\theta}_{1}^\prime,...,\boldsymbol{\theta}_{K}^\prime$ can either be accepted 
 (i.e., $[\boldsymbol{\theta}_{1}^{i+1},...,\boldsymbol{\theta}_{K}^{i+1}] \leftarrow [\boldsymbol{\theta}_{1}^\prime,...,\boldsymbol{\theta}_{K}^\prime]$)
  or rejected 
  (i.e., $[\boldsymbol{\theta}_{1}^{i+1},...,\boldsymbol{\theta}_{K}^{i+1}] \leftarrow [\boldsymbol{\theta}_{1}^i,...,\boldsymbol{\theta}_{K}^i]$).
The acceptance or rejection of the proposed scenarios
   is stochastic, with probability $\rho_i$ (defined below), 
   which is a function of $\hat{\mathcal{L}}(\boldsymbol{\theta}_{1}^\prime,...,\boldsymbol{\theta}_{K}^\prime)$ and $\hat{\mathcal{L}}(\boldsymbol{\theta}_{1}^i,...,\boldsymbol{\theta}_{K}^i)$. 

The acceptance probability $\rho_i$ is equal to $1$ when
 $\hat{\mathcal{L}}(\boldsymbol{\theta}_{1}^{\prime},...,\boldsymbol{\theta}_{K}^{\prime}) < \hat{\mathcal{L}}(\boldsymbol{\theta}_{1}^i,...,\boldsymbol{\theta}_{K}^i)$.
 That is, if the proposed scenarios decrease the current loss value,  then the proposed scenarios are accepted.
   If instead 
   $\hat{\mathcal{L}}(\boldsymbol{\theta}_{1}^{\prime},...,\boldsymbol{\theta}_{K}^{\prime}) \geq \hat{\mathcal{L}}(\boldsymbol{\theta}_{1}^i,...,\boldsymbol{\theta}_{K}^i)$, 
   then $\rho_i$  is equal to
   $$ \exp\left(\frac{ \hat{\mathcal{L}}(\boldsymbol{\theta}_{1}^i,...,\boldsymbol{\theta}_{K}^i) - \hat{\mathcal{L}}(\boldsymbol{\theta}_{1}^{\prime},...,\boldsymbol{\theta}_{K}^{\prime})}{T_i}\right),$$ where $T_i,$ $0 \leq i \leq I$, is a decreasing sequence of positive real numbers often called the ``cooling schedule'' of the algorithm. 
A common cooling schedule is $T_i = T_0\cdot r^{i-1}$, 
where $T_0$ is a constant and 
$r \in (0,1)$ is a multiplicative contraction, but other forms are  possible \citep{spall2005introduction}. In our applications, we use a piecewise-constant cooling schedule \citep{husmann2017r}. 

After simulating the outlined Markov Chain for a fixed number $I$ of iterations, the final set of scenarios $\{\boldsymbol{\theta}_{1}^{I+1},...,\boldsymbol{\theta}_{K}^{I+1}\}$  approximately minimizes the loss function $\hat{\mathcal{L}}$ \citep{belisle1992convergence}. In our ROSA implementation, we use multiple independent replicates of  Algorithm 3, 
with different initial scenarios $\boldsymbol{\theta}_1^1,...,\boldsymbol{\theta}_K^1$,
to investigate convergence of the Markov chain.
 Intuitively, if the independent chains converge,  then
 the corresponding loss values of the approximate optima
  $\hat{\mathcal{L}}(\boldsymbol{\theta}_{1}^{I+1},...,\boldsymbol{\theta}_{K}^{I+1})$ should be  nearly identical.

\vspace{\baselineskip}
\begin{algorithm}[H]
 Initialize the values of $\boldsymbol{\theta}_{1}^1,...,\boldsymbol{\theta}_{K}^1$, e.g., by sampling from a distribution over $\boldsymbol{\Theta}$ \\
 Best proposal $ \leftarrow \boldsymbol{\theta}_{1}^1,...,\boldsymbol{\theta}_{K}^1$ \\
 \For{$i=1$ to $I$}{
  New proposal $\boldsymbol{\theta}_{1}^{\prime},...,\boldsymbol{\theta}_{K}^\prime \leftarrow \textrm{Perturb}(\boldsymbol{\theta}_{1}^{i},...,\boldsymbol{\theta}_{K}^{i})$  \\
   \uIf{$\hat{\mathcal{L}}(\boldsymbol{\theta}_{1}^\prime,...,\boldsymbol{\theta}_{K}^\prime) \leq \hat{\mathcal{L}}(\boldsymbol{\theta}_{1}^{i},...,\boldsymbol{\theta}_{K}^{i})$ }
   {
    Define $\boldsymbol{\theta}_{j}^{i+1} = \boldsymbol{\theta}_{j}^\prime$ for every $j=1,\ldots, K$ \;
   }
  \uElse{ 
  Compute the acceptance probability $\rho_i =\exp\left([ \hat{\mathcal{L}}(\boldsymbol{\theta}_{1}^i,...,\boldsymbol{\theta}_{K}^i) - \hat{\mathcal{L}}(\boldsymbol{\theta}_{1}^{\prime},...,\boldsymbol{\theta}_{K}^{\prime})]/T_i\right)$ \\
  Sample $U_i\sim\textrm{Uniform}(0,1)$ \\
  If $U_i \leq \rho_i$, define $\boldsymbol{\theta}_{j}^{i+1} = \boldsymbol{\theta}_{j}^\prime$ for every $j=1,\ldots, K$; \\ Otherwise $\boldsymbol{\theta}_{j}^{i+1} = \boldsymbol{\theta}_{j}^i$ for every $j=1,\ldots, K$.
  }
 }
 \textbf{Output:} {$\boldsymbol{\theta}_{1}^{I+1},...,\boldsymbol{\theta}_{K}^{I+1}$, $\hat{\mathcal{L}}(\boldsymbol{\theta}_{1}^{I+1},...,\boldsymbol{\theta}_{K}^{I+1})$}
 \caption{Pseudocode for SA to obtain ROSA scenarios}
\end{algorithm}

\section{Applications: Sensitivity Analyses of Trial Designs}
We illustrate the ROSA approach by performing sensitivity analyses for three designs of different complexity levels. In each example, we describe the design of the trial, the UPs, and the OCs of interest. 

In the first example, we will only consider a single UP (i.e., $\boldsymbol{\theta} \in \mathbb{R}$) and a single OC $f(\boldsymbol{\theta})$ that can be computed analytically. In this case, the optimal set of scenarios $\{\boldsymbol{\theta}_1^*,...,\boldsymbol{\theta}_K^*\}$ can be computed exactly, without resorting to approximation methods. This simple and stylized setting is useful to highlight the similarity of the approximations and selected scenarios computed by ROSA with their exact counterparts.

In the second example, we consider sensitivity analyses with multiple UPs and two OCs. We illustrate the use of our computational procedures, including the OCs approximation procedure (Algorithm 1), the validation procedure (Algorithm 2), and the SA optimization procedure (Algorithm 3). We investigate whether it is appropriate to fix the value of some of the UPs across all sensitivity scenarios. Identical values for a subset of the UPs can simplify the interpretation of the sensitivity analysis but can also introduce severe limitations in faithfully representing how the OCs vary across plausible values of the UPs. 

In the third example, we discuss sensitivity analyses dedicated to an adaptive trial with sub-populations defined by biomarkers, considering multiple UPs and multiple OCs of interest. We examine the difference in the marginal losses
\begin{equation}
    \label{eq:marginal}
    \mathcal{L}_r(\boldsymbol{\theta}_1,...,\boldsymbol{\theta}_K) = \max_{\boldsymbol{\theta} \in \boldsymbol{\Theta}} \left\{\min_{k=1,...,K} ||f_r(\boldsymbol{\theta}) - f_r(\boldsymbol{\theta}_k)||_2 \right\}, \quad{} 1 \leq r \leq R,
\end{equation}
when the set of scenarios are chosen by optimizing different loss functions. For example, let $\mathcal{S}_r$ be the set of scenarios that minimize the marginal loss $\mathcal{L}_r$  in (\ref{eq:marginal}). Similarly, let $\mathcal{S}$ be the set of scenarios that minimize the joint loss $\mathcal{L} = -\mathcal{U}$  in (\ref{eq:loss}). Then it is intuitive that $\mathcal{L}_r(\mathcal{S}_r) \leq \mathcal{L}_r(\mathcal{S})$, $1 \leq r \leq R$. In different words, the marginal losses $\mathcal{L}_r$ tend to be smaller when the set of scenarios is chosen to minimize  $\mathcal{L}_r$  compared to 
a set of scenarios that minimizes $\mathcal{L}$  with the aim of representing multiple OCs. If the discrepancy  $\mathcal{L}_r(\mathcal{S}_r) - \mathcal{L}_r(\mathcal{S})$, $1 \leq r \leq R$, is relatively small for all $R$ total OCs, then this indicates that it is reasonable to select a single set of scenarios $\mathcal{S}$  to  illustrate how the  $R$ OCs vary jointly across $\boldsymbol{\Theta}$.

By illustrating the ROSA methodology in three trial designs, we show its flexibility  with potential applications  to evaluate nearly any clinical trial design. Indeed, ROSA only requires the possibility of simulating the trials under potential UPs $\boldsymbol{\theta} \in \boldsymbol{\Theta} =\mathbb{R}^{d}$ and the definition of the OCs of interest.

\subsection{Application 1: Two-arm RCT}
We consider the design of a two-arm randomized trial (1:1 randomization ratio) with a sample of $n=30$ patients. For each $i=1,\ldots,n$, we let $A_i=0$ or $1$ if the $i$-th study patient is assigned to the control or experimental arm. The outcomes of the $n$ study patients are $Y_1,\ldots,Y_n$, which we assume to be independent and normally distributed. If $A_i = a$ then $Y_i$ has mean $\boldsymbol{\mu}_a = 100 + 15a$ and standard deviation $\sigma$ equal to $30$. In the analysis of the study, a $z$-statistic will be used to test the null hypothesis $H_0: \boldsymbol{\mu}_1 - \boldsymbol{\mu}_0 \leq 0$ against the alternative $H_1: \boldsymbol{\mu}_1 - \boldsymbol{\mu}_0 > 0$ at  5\% significance level.

The goal of the sensitivity analysis is to assesses the variation of the probability of rejecting $H_0$, a function $f(\boldsymbol{\theta})$ of the unknown treatment effect $\boldsymbol{\theta} = \boldsymbol{\mu}_1 - \boldsymbol{\mu}_0\in \boldsymbol{\Theta} = \mathbb{R}$. For example, if we knew that $\boldsymbol{\theta} = 13.5$, then $f(\boldsymbol{\theta})=0.80$, but in general $\boldsymbol{\theta}$ is an unknown value. %The variations in $f({\theta})$, and so in the power or type 1 error rate of the design, can be large across potential values of ${\theta}$.

Suppose we aim to identify $K=3$ scenarios $\boldsymbol{\theta}_{1}^*,\boldsymbol{\theta}_{2}^*,\boldsymbol{\theta}_{3}^*$ that maximize the utility $\mathcal{U}$, i.e.,
\begin{equation}
    \boldsymbol{\theta}_{1}^*,\boldsymbol{\theta}_{2}^*,\boldsymbol{\theta}_{3}^* = \argmax_{\boldsymbol{\theta}_1,\boldsymbol{\theta}_2,\boldsymbol{\theta}_3\in \boldsymbol{\Theta}} \text{ } \mathcal{U}(\boldsymbol{\theta}_1,\boldsymbol{\theta}_2,\boldsymbol{\theta}_3),
\end{equation}
where $\mathcal{U}(\boldsymbol{\theta}_1,\boldsymbol{\theta}_2,\boldsymbol{\theta}_3) = -\max_{\boldsymbol{\theta}^\prime \in \boldsymbol{\Theta}} \min_{k=1,2,3} |f(\boldsymbol{\theta}^\prime) - f(\boldsymbol{\theta}_{k})|.$

\begin{figure}
    \centering
    \includegraphics[scale=0.6]{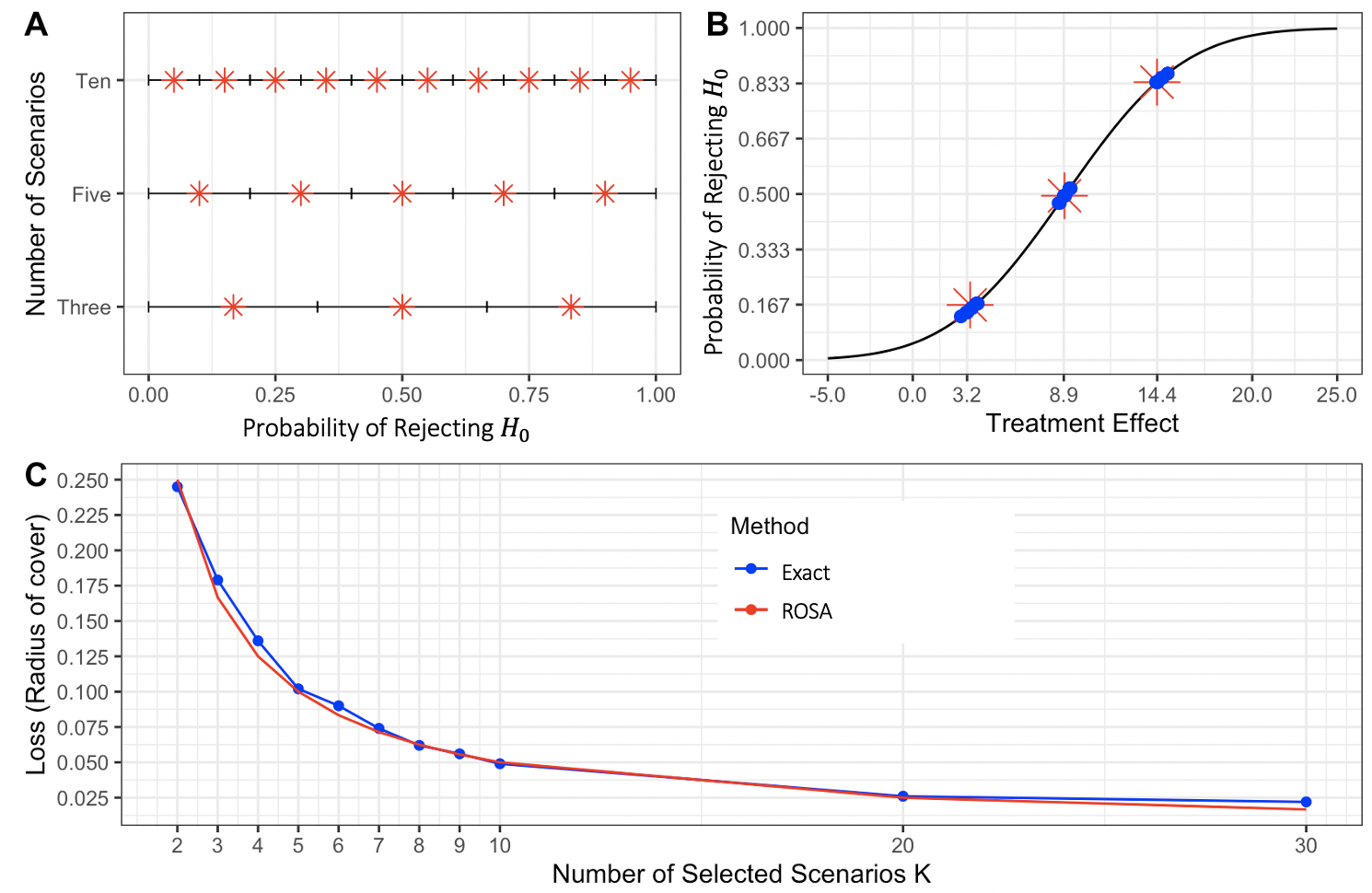}
    \caption{Sensitivity analysis of a RCT (OC: probability of rejecting $H_0$). \textbf{Panel A:}  Exact solutions when $K = \{3,5, 10\}$.  \textbf{Panel B:} Comparison of $K = 3$  scenarios selected through exact calculation {\color{red}(red asterisks)} and by $20$ ROSA implementations with different initial proposals {\color{blue}(blue points)}.  \textbf{Panel C:} Graphical tool to choose the number $K$ of sensitivity scenarios. }
    % The estimated power matches closely with the exact power curve.}
    \label{fig:app1}
\end{figure}

In this trial, we have a single UP (${\boldsymbol\Theta} = \mathbb{R}$), and the OC of interest is monotone, continuous, invertible, and ranges from $0$ to $1$. Therefore, it is straightforward to see that the optimal scenarios $\boldsymbol{\theta}_1^*,\boldsymbol{\theta}_2^*,\boldsymbol{\theta}_3^*$ correspond to the OC values that evenly divide the 
interval $(0,1)$. To be precise, $\{f(\boldsymbol{\theta}_1^*),f(\boldsymbol{\theta}_2^*),f(\boldsymbol{\theta}_3^*)\} = \{1/6, 3/6, 5/6\}$;  these are the three values of a regular grid on the interval $(0,1)$. Figure \ref{fig:app1}A illustrates the optimal set of scenarios when $K = \{3,5,10\}$. Since $f(\boldsymbol{\theta)}$ can be calculated exactly, the optimal scenarios 
$\boldsymbol{\theta}_1^*, \boldsymbol{\theta}_2^*,\boldsymbol{\theta}_3^*$ can be obtained by computing the inverse function $f^{-1}$ at the values $1/6, 3/6,$ and $5/6$. Specifically,
$$\left\{\boldsymbol{\theta}_1^*,\boldsymbol{\theta}_2^*,\boldsymbol{\theta}_3^* \right\}=\left\{  \frac{\sigma(z_{f(\boldsymbol{\theta}_1^*)}+z_{1-\alpha/2})}{\sqrt{n}}, \frac{\sigma(z_{f(\boldsymbol{\theta}_2^*)}+z_{1-\alpha/2})}{\sqrt{n}}, \frac{\sigma(z_{f(\boldsymbol{\theta}_3^*)}+z_{1-\alpha/2})}{\sqrt{n}} \right\},$$
where
$z_{1-\alpha/2}$ is the $1-\alpha/2$ quantile of the standard normal distribution. The corresponding optimal scenarios are illustrated as red asterisks in Figure \ref{fig:app1}B. 

The exact computation of the optimal set of scenarios provides a solid benchmark for an initial evaluation of ROSA (Algorthms 1-3). We can compare the exact solution with the results from ROSA, which has the advantage of being applicable to other designs and OCs that are not available in closed form. 

We implement our ROSA approach to identify $K=3$ scenarios. We randomly select $J = 1000$ scenarios $\boldsymbol{\theta}_1^t,...,\boldsymbol{\theta}_{1000}^t$ with independent samples from the $\textrm{Uniform}(-5,25)$ distribution. Note that $f(-5) \approx 0$ and $f(25) \approx 1$. For each $\boldsymbol{\theta}_j^t$, $1 \leq j \leq 1000$, we simulate $M = 200$ trials to compute the estimate $\bar{f}(\boldsymbol{\theta}_j^t) = 200^{-1}\sum_{m=1}^{200} {\varphi}(Z_{j,m}, \boldsymbol{\theta}_{j}^t),$ 
where ${\varphi}(Z_{j,m}, \boldsymbol{\theta}_{j}^t) \in \{0,1\}$ either accepts or rejects the $H_0: \boldsymbol{\theta}_j^t \leq 0$ for trial $m$ and scenario $j$. Then, we compute a smooth function $\hat{f}(\boldsymbol{\theta})$ using the independent estimates $\bar{f}(\boldsymbol{\theta}_j)$ and a NN with 3 hidden layers (8, 64, and 64 neurons respectively) and ReLU activation functions. Finally, to select three sensitivity scenarios, we use a SA algorithm based on an initial parameterization $T_1 = 1000$, temperature reduction factor $r = 0.8$, and final parameterization $T_{\min} = 0.1$ (c.f. Algorithm 3). We repeat these three steps (selection of scenarios, use of the NN, and optimization with SA) $20$ times, each time initializing $\boldsymbol{\theta}_1, \boldsymbol{\theta}_2, \boldsymbol{\theta}_3$ with independent random draws from the $\textrm{Uniform}(-5,25)$ distribution. 
% Convergence was assessed by examining trace plots; c.f. Supplementary Figure 1.
The results of the exact approach (red asterisks) compared with ROSA (blue points) are shown in Figure \ref{fig:app1}B.  The scenarios $\boldsymbol{\theta}_{1}^*,\boldsymbol{\theta}_{2}^*,\boldsymbol{\theta}_{3}^*$ selected by SA (blue dots) are close to the exact solution (red asterisks). 
% In fact, the difference in the utility $\mathcal{U}$ between the exact solution and the average of the ROSA results is less than $1\%$.

We ran ROSA with $K = 2,3,4,5,6,7,8,9,10,20,$ or $30$, and compared the loss $\mathcal{L}$ in the resulting set of scenarios with that of the exact solution. The difference in the loss $\mathcal{L}$ of the exact and approximate optima was less than $1\%$ across all $K$ values that we considered (Figure \ref{fig:app1}C). Table \ref{table1} indicates that the computation time of the SA algorithm scales well as $K$ increases and that, as expected, the loss $\mathcal{L}$ decreases as $K$ increases. All analyses were run on a Windows laptop with an Intel(R) Core(TM) i7-7700HQ 2.80 GHz processor, 16GB RAM, and 6MB of cache memory.

In practice, the decision regarding the number $K$ of scenarios to report is  left to the analyst. This choice can be supported by a graph  like Figure \ref{fig:app1}C, which allows the investigator to determine the minimum number $K$ of scenarios needed to guarantee a loss $\mathcal{L}(\boldsymbol{\theta}_1^*,...,\boldsymbol{\theta}_K^*)$ 
no larger than a targeted threshold. For example, to guarantee a loss no larger than $0.050$ in this example, we  need to select at least $10$ scenarios for the sensitivity report.

\begin{table}[H]
	\centering
	\begin{tabular}{ccccc}
		\hline
		Number $K$ of Scenarios & Time (seconds) & ROSA Loss ${\mathcal{L}}$ & Min. Loss $\mathcal{L}$ & Rel. Diff. \\
		\hline
% 		1 & 8.6 & 0.501 & 0.500 & 0.2\%  \\
% 		2 & 8.4 & 0.250 & 0.250 & 0.0\%  \\ 
% 		3 & 8.5 & 0.168 & 0.167 & 0.8\%  \\ 
% 		4 & 9.0 & 0.126 & 0.125 & 0.8\% \\ 
		5 & 8.8 & 0.101 & 0.100 & 1.0\% \\
		6 & 8.8 & 0.084 & 0.083 & 0.7\% \\ 
		7 & 9.1 & 0.072 & 0.071 & 0.8\%   \\ 
		8 & 9.2 & 0.062 & 0.0625 & 0.7\%  \\ 
		9 & 9.1 & 0.056 & 0.056 & 0.6\% \\ 
		10 & 9.1 & 0.050 & 0.050 & 0.2\% \\ 
		20 & 10.1 & 0.025 & 0.025 & 0.5\%  \\
		 30 & 10.2 & 0.017 & 0.0167 & 0.8\% \\
		\hline
	\end{tabular}
	\caption{ROSA computation time, ROSA loss $\mathcal{L}$, minimum (exact) loss $\mathcal{L}$, and relative difference in loss of ROSA scenarios compared to the exact solutions.}
	\label{table1}
\end{table}

\subsection{Application 2: Interim decisions based on auxiliary outcomes}

We consider a two-arm, two-stage randomized trial with a binary primary outcome $Y$ and a binary auxiliary outcome $S$ \citep{niewczas2019interim}. The primary outcome $Y$ is available  $T_Y$ months after randomization, while the auxiliary outcome $S$ is available after $T_S < T_Y$ months.
For example, in glioblastoma trials, 12-month progression-free survival (PFS) and 24-month overall survival (OS) have been used as auxiliary and primary outcomes, respectively \citep{han2014progression}.  The approach that we illustrate is applicable for any value of $T_Y$ and $T_S < T_Y$.

We let $N_a$ be the planned number of patients for arms $a=0,1$ 
(i.e., control and experimental arms) and indicate with $p_a$ the response probability $P(Y = 1 \mid A=a)$. Similarly, let $n_a$ be the planned number of patients assigned to arm $a$ before the interim analysis, and $q_a$ indicate the response probability $P(S=1 \mid A=a)$. 
The difference $\Delta = p_1-p_0$ is the treatment effect on $Y$. 
The primary aim of the trial is to test  $H_0: \Delta \leq 0$ versus $H_1: \Delta > 0,$ at level $\alpha$. The final analysis of the study involves only the primary outcome $Y$, and the trial will use a standard $Z$-test, $Z_Y = \frac{\hat{p}_1-\hat{p}_0}{\sqrt{\bar{p}(1-\bar{p})(N_1^{-1}+N_0^{-1})}},$ where  $\hat{p}_a$ is the estimate of $p_a$ and $\bar{p} $ is a weighted average of $\hat{p}_1 $ and $ \hat{p}_0$.

An interim analysis is conducted after the auxiliary outcomes $S$ become available for $n_a$ patients for arms $a = 0$ and $1$ (i.e., $T_S$ months after the 
 enrollment of  $n_a$ patients on arms $a = 0 $ and $1$), with early-stopping for futility or continuation based on a  summary of the auxiliary outcomes $S$. In several clinical settings, the treatment effect on $S$ tends to be more pronounced than the treatment effect on $Y$. The interim analysis is based on the summary $Z_S = \frac{\hat{q}_1-\hat{q}_0}{\sqrt{\bar{q}(1-\bar{q})(n_1^{-1}+n_0^{-1})}},$ where $\hat{q}_a$ is the estimate of $q_a$ and $\bar{q} $ is a weighted average  of $\hat{q}_1 $ and $\hat{q}_0$.
 We replicate the design of \cite{niewczas2019interim}, which calculates at the interim analysis the conditional power (CP) using the auxiliary outcome $S$ to determine whether to stop the trial for futility or not. Specifically, the CP is calculated based on $Z_S$ and the information fraction $t_S = \frac{N_1^{-1} + N_0^{-1}}{n_1^{-1} + n_0^{-1}}$ as
$$CP(t_S) = 1- \Phi\left(\frac{ z_{1-\alpha} - Z_{S} t_S^{1/2} }{\sqrt{1-t_S}} \right),$$
where $z_{1-\alpha}$ is the $1-\alpha$ quantile of the standard normal distribution and $\Phi(\cdot)$ is the cumulative distribution function of the standard normal distribution.
Here, we set the cut-off point to be $0.5$ so that the trial continues when $CP(t_s) \geq 0.5$. 

The complexity of the sensitivity report increases with 
$K$ (the number of scenarios), 
$d$ (the number of entries of the UPs $\boldsymbol{\theta}$), 
and $R$ (the number of OCs $\textbf{f}(\boldsymbol{\theta})$). 
 Here the full set of UPs $\boldsymbol{\Theta} \subset \mathbb{R}^7$
  include the enrollment rate $e \in (0,\infty)$, 
  the response rates $p_a \in (0,1)$ for $Y$ in $A=a$, 
  the response rates $q_a \in (0,1)$ for $S$ in $A=a$, 
  and the correlation between $Y$ and $S$ in $A=a$, $\rho_a \in (-1,1)$. 
  
 Controlling the complexity of the sensitivity report is important  to ensure high interpretability of the  report, which will be discussed by several stakeholders. 
  There are a few potential strategies to reduce the complexity of the sensitivity report. First, it is often possible to consider only a subset of the parameter space $\boldsymbol{\Theta}^\prime \subset \boldsymbol{\Theta}$ based on prior knowledge of plausible values of the UPs. For example, previous clinical studies can indicate a plausible range for the enrollment rate $e$, the response rates $p_0$ under the SOC, and other parameters that are expected to have  minimal variations across trials. In addition, we can also consider fixing multiple entries of the $K$ vectors $\boldsymbol{\theta}_1,...,\boldsymbol{\theta}_K$ to some reference values.
  In this case
the  space  from which we select 
scenarios $\boldsymbol{\theta}_1,...,\boldsymbol{\theta}_K$
is further reduced to $\boldsymbol{\Theta}^\prime_{re} \subset \boldsymbol{\Theta}^\prime$. 
 For example, if the OCs have low sensitivity with respect to 
 the correlation parameters $\rho_a$ or
 the enrollment rate $e$ of the study, then we can fix these UPs to common values (i.e., estimates) across all $K$ scenarios. 
 
 ROSA allows us to evaluate whether it is appropriate to assign the same value to one or more UPs (e.g., $\rho_0$ and $\rho_1$) across all $K$ scenarios. In other words, we  evaluate a sensitivity report with all scenarios in a   restricted subset $\boldsymbol{\Theta}^\prime_{re} \subset \boldsymbol{\Theta}^\prime$. A sensitivity report with scenarios in   $\boldsymbol{\Theta}^\prime_{re}$ can potentially
be easier to interpret  compared to a report in which all $d$ entries of $\boldsymbol{\theta}$ vary across scenarios by reducing  the number of dimensions $d$ of the UPs and pointing to the most relevant UPs  when discussing  the variations of the OCs across  $\boldsymbol{\Theta}^\prime$. We can select scenarios from the restriction  $\boldsymbol{\Theta}_{re}^\prime \subset \boldsymbol{\Theta}^\prime$   only if the capability of the sensitivity report of representing the OCs variations across $\boldsymbol{\Theta}^\prime$ is preserved. Our  case study investigates this aspect.

The OCs of interest in our case study are the probability of rejecting the null hypothesis of no treatment effect on $Y$ at the end of the study and the average sample size.
 Using our procedure, we randomly select $J = 1000$ training scenarios using LHS and conduct $M = 500$ Monte Carlo simulations for each of the $J$ training scenarios to obtain estimates of the OCs across $\boldsymbol{\Theta}^\prime$. %We then repeat the procedure to obtain estimates of the OCs across the restriction $\boldsymbol{\Theta}^\prime_{re}$. .
 
 Here $\boldsymbol{\Theta}^\prime$ is a product space with the enrollment rate $e \in (0.2,1)$, the response rates $p_a \in (0.2,0.4)$ for $Y$ in $A=a$, the response rates $q_a \in (0.2,0.4)$ for $S$ in $A=a$, and the correlation between $Y$ and $S$ in $A=a$, $\rho_a \in (0,0.6)$. For $\boldsymbol{\Theta}^\prime_{re}$, we fix the enrollment rate $e = 0.5$ and the response rates $p_0 = q_0 = 0.3$ in the control groups.

   We use a NN to obtain an interpolation of the OCs. As described in Algorithm 4, to evaluate if the estimates of the OCs are accurate, we compare them to independent Monte Carlo estimates of size M = 100,000 on a set of $J^\prime = 200$ uniformly-distributed validation points spanning the plausible parameter space $\boldsymbol{\Theta}^\prime$. The coefficient of determination $R^2$ in this comparison is $0.96$, suggesting that the NN accurately estimates the OCs (Supplementary Materials).

\begin{figure}[H]
    \centering
    \includegraphics[scale=0.52]{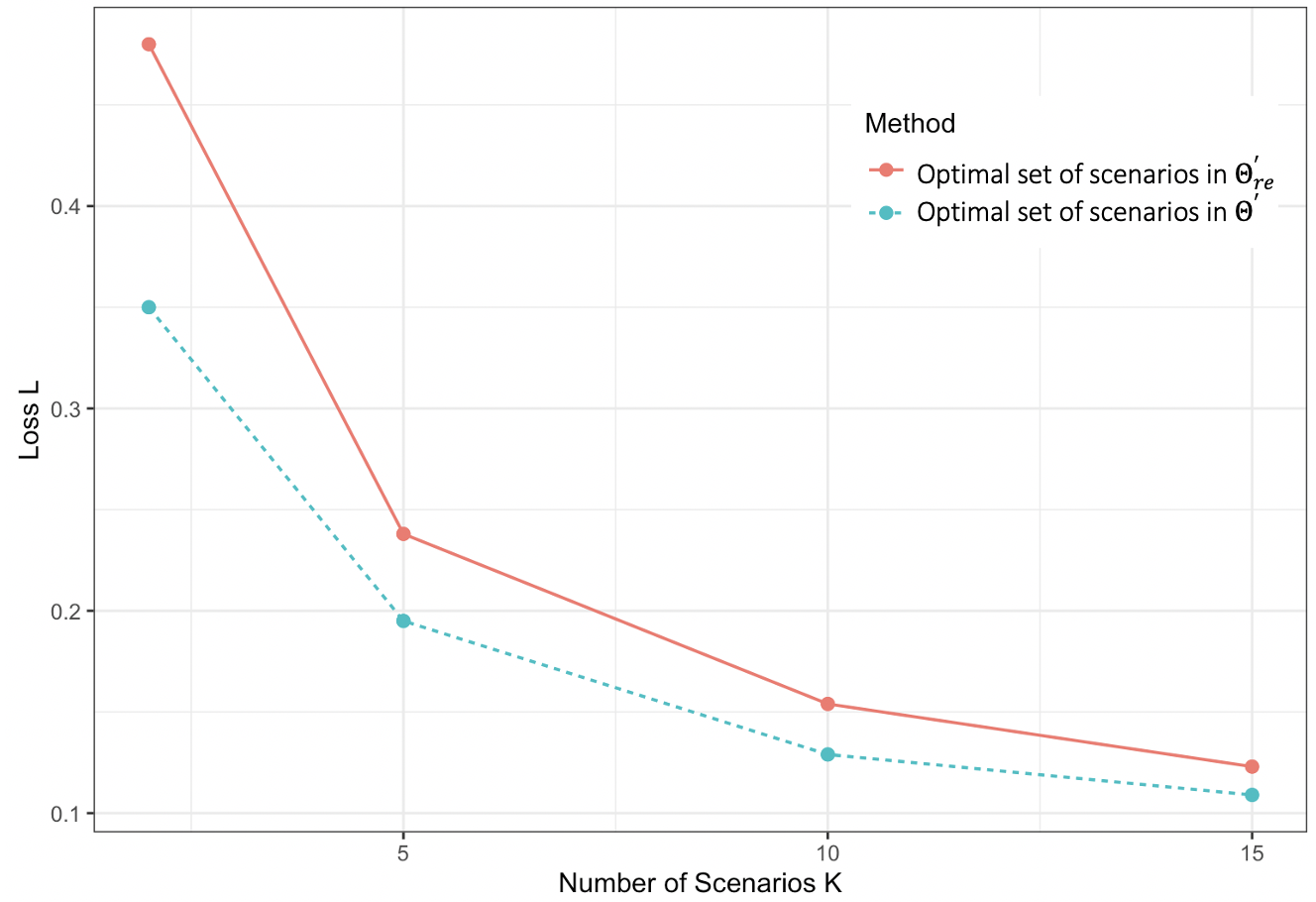}
    \caption{Clinical trial design with an interim analysis and an auxiliary endpoint. A graphical representation to choose the number of sensitivity scenarios $K \in \{2,5,10,15\}$. 
   We compare optimal sets of scenarios selected from
    $\boldsymbol{\Theta}^\prime \subset \mathbb{R}^7$ and from the
    lower-dimensional 
     restriction  $\boldsymbol{\Theta}^\prime_{re} \subset \boldsymbol{\Theta}^\prime$.  }
    \label{fig:app2}
\end{figure}

We compare two sensitivity reports, and our goal is to provide stakeholders the simplified version if it accurately describes the OCs. The first one includes scenarios from $\boldsymbol{\Theta}^\prime \subset \mathbb{R}^7$ restricted by prior knowledge from completed studies and clinical experience and the second  includes scenarios from $\boldsymbol{\Theta}^\prime_{re} \subset \boldsymbol{\Theta}^\prime $ further restricted by fixing the value of some entries of $\boldsymbol{\theta}$ as described above. We use SA to identify two sets of  scenarios in $\boldsymbol{\Theta}^\prime_{re} $ and $ \boldsymbol{\Theta}^\prime $, respectively.
In both cases we minimize the same loss function $\mathcal{L}$ defined over K-tuples of $ \boldsymbol{\Theta}^\prime $ points.
We also calculate the  loss $\mathcal{L}$ associated with these two  optimal sets of   scenarios from $\boldsymbol{\Theta}^\prime$ and $\boldsymbol{\Theta}^\prime_{re}$.
In Figure \ref{fig:app2}, we illustrate the difference in  loss $\mathcal{L}$ between these two optimal sets. As expected, the loss $\mathcal{L}$ decreases as $K$ increases. 
We observe in Figure \ref{fig:app2}  that for any  value of $K$, the loss $\mathcal{L}$ associated with 
the optimal set of scenarios restricted to $\boldsymbol{\Theta}^\prime_{re}$ 
is larger compared to 
the optimal scenarios in $\boldsymbol{\Theta}^\prime$.
However, the difference is modest, and the gain in interpretability of a sensitivity analysis report with fewer UPs may be worth the slightly larger loss. For example, if an investigator requires the loss to be under a threshold of $\mathcal{L} = 0.2$, then it is sufficient to consider $K=10$ scenarios, regardless of whether we consider scenarios selected from $\boldsymbol{\Theta}^\prime$ or $\boldsymbol{\Theta}^\prime_{re}$.

\subsection{Application 3: Biomarker-driven adaptive enrichment}
In several oncology trials, a major decision is whether to restrict patient enrollment to a targeted subgroup of patients (e.g., biomarker-positive subgroup) or to enroll a broader patient population. Enrolling only a biomarker-positive subgroup may deny a substantial number of patients access to an effective therapy, whereas enrolling a larger population may compromise the power to detect positive treatment effects. Several trial designs discussed in the literature attempt to address the outlined problem through interim looks at the data. Among them, we consider an adaptive two-stage enrichment trial design with one-to-one randomization \citep{jenkins2011adaptive, jones2017tappas, mehta2019adaptive}. The design is applicable in the setting where a biomarker-positive subgroup of patients is hypothesized to benefit more from the experimental treatment than the rest of the study population.

The design includes a single interim analysis, and it uses progression-free survival (PFS) for interim decision-making, while overall survival (OS) is the endpoint for the final analysis, which occurs when a pre-specified number of events is reached. The interim analysis uses the estimated PFS hazard ratio (HR) to capture potential early signals of treatment effects. In the implementation of \cite{jenkins2011adaptive}, which we replicate, the HR is estimated for both the overall population ($\hat{\theta}_{HR}$) and the biomarker-positive subgroup ($\hat{\theta}_{HR}^+$). An interim decision determines which group is enrolled and tested during the second stage of the trial: 

A -- {\tt Promising results in the biomarker-positive population.}
If the HR estimate $\hat{\theta}_{HR}^+ < 0.6$ but  $\hat{\theta}_{HR} \geq 0.8$, then the trial will continue enrolling only biomarker-positive patients and the final analysis will test $H_0^+$. Here $H_0^+$ is the null hypothesis of no differences in OS between treatment and control groups in the biomarker-positive population. The null hypothesis is rejected if $ \omega_1 \Phi^{-1}(1-p_1^+) + \omega_2 \Phi^{-1}(1-p_2^+) < 1.96$, where $p_1^+$  $(p_2^+)$ is a log-rank p-value computed using  only OS data from patients randomized during the first (second) stage of the trial.  The weights $(\omega_1,\omega_2)$ and the standard normal cumulative distribution function  $\Phi$ are used to summarize evidence of treatment effects from the two stages of the trial. We refer to \cite{jenkins2011adaptive} for details on the choice of $(\omega_1,\omega_2)$ and other aspects of the final analysis. 
 
B -- {\tt Promising results in the overall population only.}
 If $\hat{\theta}_{HR}^+ \geq 0.6$ but  $\hat{\theta}_{HR} < 0.8$, then the trial will continue enrolling all patients and the final analysis will only test $H_0^O$, the null  hypothesis of no differences in OS  in the overall population. In this case the null hypothesis is tested using stage-specific OS log-rank p-values $(p_1^O,p_2^O)$ and combining evidence from the two stages of the trial.

C -- {\tt Unpromising results.}  If $\hat{\theta}_{HR}^+ \geq 0.6$ and  $\hat{\theta}_{HR} \geq 0.8$, then the trial  stops early for futility.

D -- {\tt Promising early results for both populations.} Lastly, if the estimated HR in the biomarker-positive subgroup $\hat{\theta}_{HR}^+ < 0.6$ and the overall population $\hat{\theta}_{HR} < 0.8$, then the trial will continue enrolling all patients and testing efficacy  both in the overall population and in the biomarker-positive subgroup. 

The potential conclusion at the final analysis are (i) to recommend the new treatment for biomarker-positive patients, (ii) recommend the new treatment for both biomarker-positive and biomarker-negative patients, or (iii) not recommend the experimental treatment for future patients.

{\it Sensitivity analysis; the definition of  $\boldsymbol{\Theta}^\prime$.}
We choose plausible intervals for the UPs based on prior literature.
Specifically, the recruitment rate $\theta_1 \in (0.5,1)$ per week,
the prevalence of the biomarker-positive subgroup $\theta_2 \in (0.15,0.25)$,
the PFS HR comparing the treatment and control groups in the biomarker-positive subgroup $\theta_3 \in (0.5,1.2)$, 
the PFS HR comparing  treatment and control in the biomarker-negative subgroup $\theta_4 \in (0.6,1.2),$ 
the OS HR comparing  treatment and control  in the biomarker-positive subgroup $\theta_5 \in (0.7,1.2),$
the OS HR comparing  treatment and control groups in the biomarker-negative subgroup $\theta_6 \in (0.8,1.2),$
the correlation between OS and PFS in the biomarker-positive subgroup $\theta_7 \in (0.3,0.6)$,
and the correlation between OS and PFS in the biomarker-negative subgroup $\theta_{8} \in (0.2,0.7)$.
Marginal exponential distributions using a mixture representation were used for simulating correlated OS and PFS times \citep{michael2002}. More flexible models such as the Weibull distribution can be considered.

%%%%%%%%%%%%%%%%%%%%%%%%%%%%

%%%%%%%%%%%%%%%%%%%%%%%%%%%%
%%%%%%%%%%%%%%%%%%%%%%%%%%%%

%%%%%%%%%%%%%%%%%%%%%%%%%%%%
%%%%%%%%%%%%%%%%%%%%%%%%%%%%

%%%%%%%%%%%%%%%%%%%%%%%%%%%%
%%%%%%%%%%%%%%%%%%%%%%%%%%%%

%%%%%%%%%%%%%%%%%%%%%%%%%%%%
%%%%%%%%%%%%%%%%%%%%%%%%%%%%

{\it Sensitivity analysis; the definition of  $\boldsymbol{f}$.}
We focus on the following three OCs:
(i) $f_1$, the probability of enrolling only biomarker-positive patients  in the second stage,
(ii) $f_2$, the probability of enrolling both biomarker-positive and biomarker-negative patients in the second stage,
and (iii) $f_3$, the probability of no evidence of positive treatment effects, which is equal to the probability of not rejecting the null hypotheses.

For the outlined two-stage trial with biomarker populations, our ROSA pipeline can be used to compute multiple sensitivity reports, varying both the list of OCs $\boldsymbol{f}$ and the definition of $\boldsymbol{\Theta}'$. For example, one can fix the OS HRs in the biomarker-positive  and negative populations to focus on the  design sensitivity to other parameters, such as the PFS HRs. Similarly, the set of UPs $\boldsymbol{\Theta}'$ can be restricted to $\boldsymbol{\theta}$ values with positive effects only for the biomarker-positive  population.
Importantly, one set of training simulations can be re-utilized to compute multiple sensitivity tables where the definitions of $\boldsymbol{f}$ and $\boldsymbol{\Theta}'$ vary.

\begin{figure}[H]
    \centering
    \includegraphics[scale=0.18]{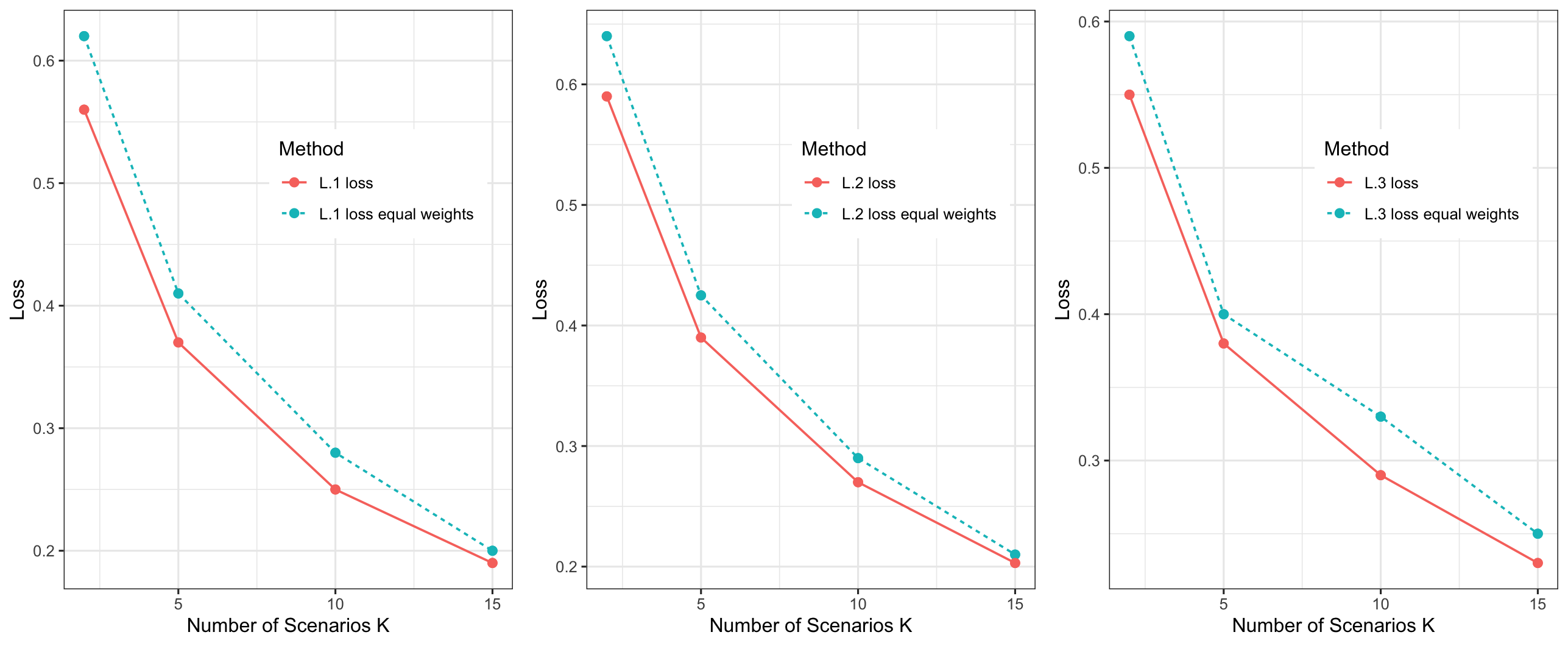}
    \caption{Marginal losses $\mathcal{L}_r$, $r=1,2,3$ of different sets of scenarios $\mathcal{S}_r$ (red) and $\mathcal{S}$ (blue).}
    \label{fig:app3}
\end{figure}

We describe the difference between the marginal losses $\mathcal{L}_r$, $r=1,2,3$, when  scenarios $ \boldsymbol{\theta}_1,\ldots, \boldsymbol{\theta}_K $ in $\boldsymbol{\Theta}'$  are chosen by optimizing $\mathcal{L}_r$  in (\ref{eq:marginal}) -- optimum: $\mathcal{S}_r = \boldsymbol{\theta}_{1,r}^*,...,\boldsymbol{\theta}_{K,r}^*$ -- or by optimizing $\mathcal{L}$ as in (\ref{eq:loss}) -- optimum: $\mathcal{S} = \boldsymbol{\theta}_1^*,...,\boldsymbol{\theta}_K^*$. 
Recall that $\mathcal{S}$ is computed with the goal of illustrating how multiple OCs vary across  $\boldsymbol{\Theta}'$ while  $\mathcal{S}_r$ optimizes the representation of a single OC $f_r$.
The 
weights in (\ref{eq:loss}) are 
 $w_1=w_2=w_3 = 1/3$.  
 In Figure \ref{fig:app3} panel 1, we plot $\mathcal{L}_1(\mathcal{S}_1)$ in red and $\mathcal{L}_1(\mathcal{S})$ in blue. Similarly, in panel 2 we compare $\mathcal{L}_2(\mathcal{S}_2)$ and $\mathcal{L}_2(\mathcal{S})$, and in panel 3 we compare $\mathcal{L}_3(\mathcal{S}_3)$ and $\mathcal{L}_3(\mathcal{S})$. Our results indicate that for all three OCs, $\mathcal{L}_r(\mathcal{S}) > \mathcal{L}_r(\mathcal{S}_r)$, $r=1,2,3$. As expected,  there is an increase of the marginal losses $\mathcal{L}_r$  when the set of scenarios is selected to illustrate jointly the variations of multiple OCs  across $\boldsymbol{\Theta}'$. However, this difference  is small $(<10\%)$ for all $K \in \{2,5,10,15\}$. Furthermore, for each  $K\in \{2,5,10,15\}$, the  relative difference is  similar  across the three OCs $f_1,f_2,f_3$ (Figure \ref{fig:app3}).
This result supports the use of identical weights and of a single sensitivity table, with the same set of scenarios $\mathcal{S}$   to illustrate jointly all three OCs.

\section{Discussion}
 During the design stage of a clinical trial,  sensitivity reports are typically produced to discuss sample size, interim analyses, and other major decisions with various stakeholders. The sensitivity report consists of one or a few  tables dedicated to showcasing how major operating characteristics (OCs) $\mathbf{f}(\boldsymbol{\theta})$ vary across potential values of   UPs in $\boldsymbol{\Theta}$. In most cases the analyst  focuses on  subsets of plausible parameters $\boldsymbol{\Theta}^\prime \subset \boldsymbol{\Theta}$, for example,
 values concordant with previous studies, 
 or subsets  of potential
 $\boldsymbol{\theta}$ values of particular interest  becouse of  positive and clinically relevant treatment effects. 
ROSA  supports the choice of which and how many scenarios to include in these sensitivity reports.

The evaluation of complex designs such as dose-finding studies \citep{iasonos2015scientific}, factorial trials \citep{green2002factorial}, and response-adaptive trials \citep{pallmann2018adaptive} focuses on multiple OCs, such as the level of  toxicities, the probability of selecting the correct treatment arm, or frequentist OCs, including  power and false positive  probabilities.

Simulations are fundamental in the design of complex trials since OCs can rarely be obtained analytically and are crucial in the assessment  of study designs for regulators, pharmaceutical companies and other stakeholders \citep{food2020interacting}. 
However, a limited number of scenarios or poorly chosen scenarios  could be  inadequate to highlight variations of the OCs across plausible UPs and can result in sub-optimal decisions.  
 
We focus on choosing an informative number $K$ of scenarios $\boldsymbol{\theta}_1,...,\boldsymbol{\theta}_K$ among the plausible UPs to summarize the variations of key OCs. Our approach minimizes an explicit loss function and uses established techniques for functional approximation (NNs) and numerical optimization (SA). We showcase our approach in three trials. Importantly, our approach is general and can be applied to nearly any clinical trial  design. It only requires simulations to mimic  the clinical trial under hypothetical scenarios.  

Although our approach is general, we focused on loss functions $\mathcal{L}$ of a specific form (\ref{eq:loss}). It is possible to consider different loss functions. For example, one could consider the loss function 
$
    \tilde{\mathcal{L}}(\boldsymbol{\theta}_1,...,\boldsymbol{\theta}_K) = E_{\boldsymbol{\theta}^\prime \sim g(\cdot)}\left\{\min_{k = 1,...,K} D[\mathbf{f}(\boldsymbol{\theta}^\prime), \mathbf{f}(\boldsymbol{\theta}_k)] \right\},
$
where $g(\cdot)$ is a probability distribution on $\boldsymbol{\Theta}$ (e.g., a posterior distribution obtained from previous data). The distribution $g$ could be used to incorporate prior information about the unknown UPs in the selection of sensitivity scenarios. Moreover, the metric $D: \boldsymbol{\Theta}^2\rightarrow \mathbb{R} $ can be extended to capture both differences between OCs at plausible values $\boldsymbol{\theta}, \boldsymbol{\theta}^\prime \in \boldsymbol{\Theta}$ and other aspects, such as the difference between expected values of the outcomes $Y$ at $\boldsymbol{\theta}$ and $\boldsymbol{\theta}^\prime$.

One major challenge  in the presentation of  sensitivity reports is the need of  simplicity and interpretability of the results. To this end, we considered fixing one or more UPs   to identical values across the $K$ scenarios, which may be reasonable when there is a priori knowledge on certain UPs. There are other ways to simplify a sensitivity report, such as removing OCs that do not vary across plausible UPs, or reporting only the range of the OCs across $\boldsymbol{\Theta}$ instead of presenting the  OCs for each representative scenario.

Variations of the ROSA approach may also consider optimization algorithms other than SA and regression methods  alternative to  NN for approximating the OCs across $\boldsymbol{\Theta}$.

\section*{Acknowledgements}

The authors thank Cyrus Mehta and Christina Howe for helpful conversations and feedback that greatly enhanced the paper. LH was supported by the Clinical Orthopedic and Musculoskeletal Education and Training (COMET) Program, NIAMS grant T32 AR055885. LT was supported by NIH grant R01LM013352.  \vspace*{-8pt}

\section*{Supplementary Material}
The supplementary material includes a table of notation used in the paper,  and a validation comparison of power estimates by MC simulations and our NN in the two arm, two-stage randomized trial example.
% \begin{group}
\renewcommand{\arraystretch}{0.75}
\begin{spacing}{1.5}
\begin{table}[H]
\caption{Notation}
\small
\begin{center}
\begin{tabular}{r c p{12cm} }
\toprule
$\boldsymbol{\Theta}$ & $\triangleq$ & Unknown parameter (UP) space in $\mathbb{R}^d$ \\

$\boldsymbol{\Theta}^\prime$ & $\triangleq$ & Restricted UP subspace by prior knowledge in $\mathbb{R}^d$ \\

$\boldsymbol{\Theta}^\prime_{re}$ & $\triangleq$ & Restricted UP subspace by prior knowledge and fixing certain dimensions in $\mathbb{R}^d$ \\

$\boldsymbol{\Theta}^F$ & $\triangleq$ & Diffuse and finite UP subspace in $\mathbb{R}^d$ \\

$\boldsymbol{\theta} = (\theta_1,...,\theta_d)$ & $\triangleq$ & $d$-dimensional vector of UPs \\

$\boldsymbol{\theta}^t = (\boldsymbol{\theta}_1^t,...,\theta_d^t)$ & $\triangleq$ & $d$-dimensional training vector of UPs \\

$\boldsymbol{\theta}^v = (\theta_1^v,...,\theta_d^v)$ & $\triangleq$ & $d$-dimensional validation vector of UPs \\

$\{\boldsymbol{\theta}_1,...,\boldsymbol{\theta}_K\}$ 
& $\triangleq$ & A set of $K$ sensitivity scenarios \\

$\mathcal{S} = \{\boldsymbol{\theta}_1^*,...,\boldsymbol{\theta}_K^*\}$ 
& $\triangleq$ & The ROSA set of $K$ sensitivity scenarios optimizing loss $\mathcal{L}$ \\

$\mathcal{S}_r = \{\boldsymbol{\theta}_{1,r}^*,...,\boldsymbol{\theta}_{K,r}^*\}$ 
& $\triangleq$ & The ROSA set of $K$ sensitivity scenarios optimizing marginal loss $\mathcal{L}_r$ \\

$\mathbf{f}(\boldsymbol{\theta})$ 
& $\triangleq$ & $R$-vector of operating characteristics (OCs) for UPs $\boldsymbol{\theta}$ \\

$\hat{\mathbf{f}}(\boldsymbol{\theta})$ 
& $\triangleq$ & Estimated $R$-vector of OCs for UPs $\boldsymbol{\theta}$ \\

$\bar{\mathbf{f}}(\boldsymbol{\theta})$ 
& $\triangleq$ & Average across $M$ simulations of the $R$-vector of OCs for UPs $\boldsymbol{\theta}$ \\

$\boldsymbol{\varphi}(Z_{j,m}, \boldsymbol{\theta}_j)$ & $\triangleq$ & Generic function to capture if a null hypothesis has been rejected, where $Z_{j,m}$ is the $m^{th}$ trial under the $j^{th}$ scenario, $\boldsymbol{\theta}_j$\\

$\mathcal{L}(\boldsymbol{\theta}_1,...,\boldsymbol{\theta}_K)$ & $\triangleq$ & Loss function \\

$\mathcal{U}(\boldsymbol{\theta}_1,...,\boldsymbol{\theta}_K)$ & $\triangleq$ & Utility criterion \\

$w_1,...,w_r$ & $\triangleq$ & Fixed non-negative weights for OCs $f_1,...,f_R$ \\

$\omega_1,\omega_2$ & $\triangleq$ & Weights for stage 1 and 2 p-values \\

$D[\cdot,\cdot]$ & $\triangleq$ & Pre-specified distance metric\\

$\boldsymbol{z}_{1}^i,...,\boldsymbol{z}_{K}^i$ & $\triangleq$ & Gaussian noise in iteration $i$ of SA \\

$\rho_i$ & $\triangleq$ & Acceptance probability in iteration $i$ of SA \\

$T_0,T_1,...,T_I$ & $\triangleq$ & Decreasing sequence of positive numbers (cooling schedule of SA)\\

$r$ & $\triangleq$ & Multiplicative reduction factor for SA in $(0,1)$ \\

$U_i$ & $\triangleq$ & Random variable distributed Uniform(0,1) for SA \\

$e$ & $\triangleq$ & Enrollment rate in $(0,\infty)$ \\

$N_a$ & $\triangleq$ & Planned number of patients on arm $a = 0,1$ at the final analysis \\

$n_a$ & $\triangleq$ & Planned number of patients on arm $a = 0,1$ at the interim analysis \\

$S$ & $\triangleq$ & Binary auxiliary outcome \\

$Y$ & $\triangleq$ & Primary outcome \\

$p_a$ & $\triangleq$ & Response probability $P(Y=1 \mid A=a)$ \\

$\Delta = p_1 - p_0$ & $\triangleq$ & Treatment effect on $Y$ \\

$q_a$ & $\triangleq$ & Response probability $P(S=1 \mid A=a)$ \\

$\rho_a$ & $\triangleq$ & Correlation between $Y$ and $S$ in $A=a$ \\

\bottomrule
\end{tabular}
\end{center}
\label{tab:notation}
\end{table}
\end{spacing}

% \end{group}

% \begin{figure}[H]
%     \centering
%     \includegraphics[scale=0.55]{Old/trace2-app1.jpeg}
%     \caption{Trace plot to assess convergence for a two-arm RCT design.}
% \end{figure}

\begin{figure}[H]
    \centering
    \includegraphics[scale=0.55]{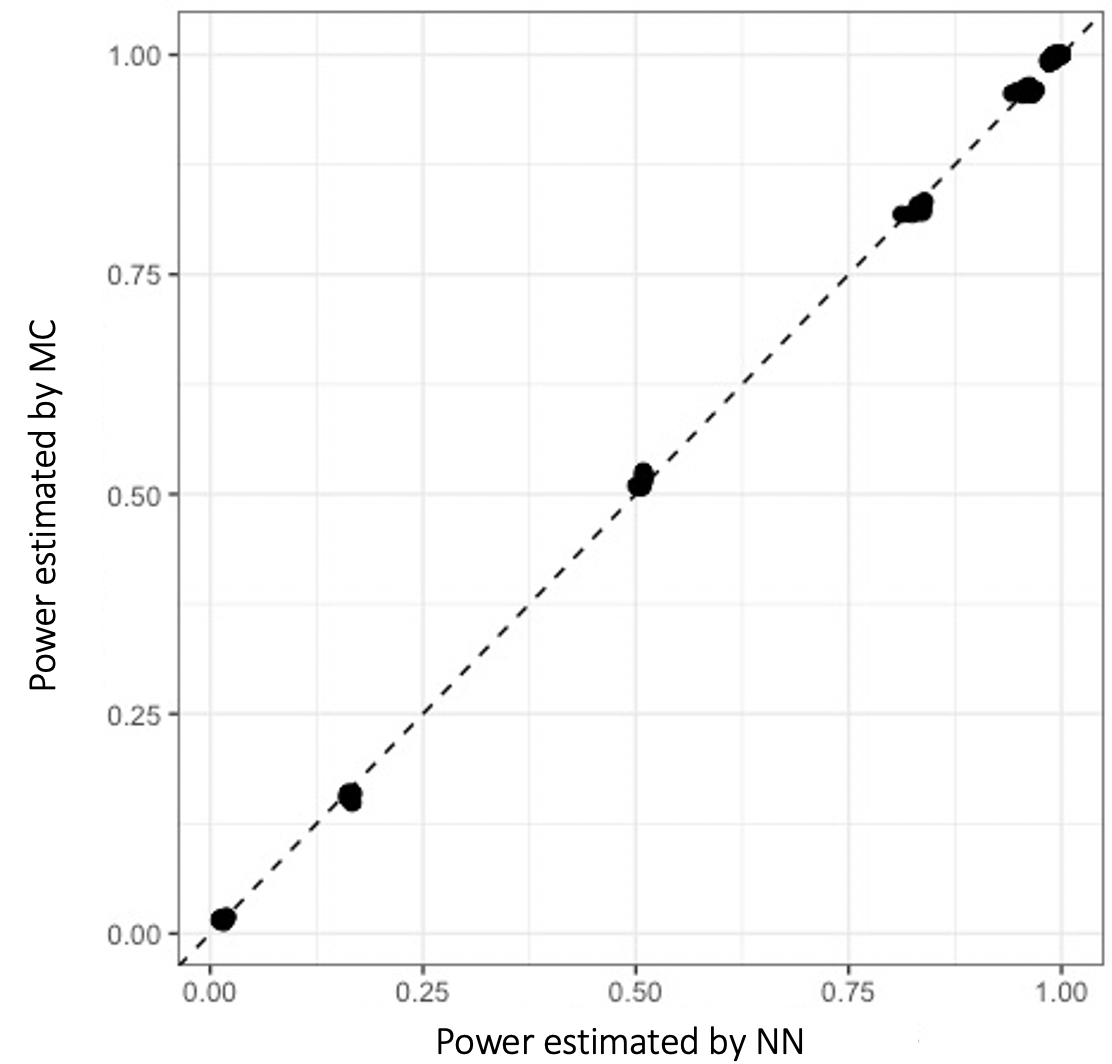}
    \caption{Comparison of power estimates from $200$ randomly sampled validation points spanning the parameter space by Monte Carlo simulation and NN.}
\end{figure}

\bibliographystyle{biom}

\bibliography{ref_LP}

\begin{thebibliography}{}

\bibitem[\protect\citeauthoryear{B{\'e}lisle}{B{\'e}lisle}{1992}]{belisle1992convergence}
B{\'e}lisle, C.~J. (1992).
\newblock Convergence theorems for a class of simulated annealing algorithms on
  rd.
\newblock {\em Journal of Applied Probability} pages 885--895.

\bibitem[\protect\citeauthoryear{Berry, Carlin, Lee, and Muller}{Berry
  et~al.}{2010}]{berry2010bayesian}
Berry, S.~M., Carlin, B.~P., Lee, J.~J., and Muller, P. (2010).
\newblock {\em Bayesian adaptive methods for clinical trials}.
\newblock CRC press.

\bibitem[\protect\citeauthoryear{Bookstein}{Bookstein}{1989}]{bookstein1989principal}
Bookstein, F.~L. (1989).
\newblock Principal warps: Thin-plate splines and the decomposition of
  deformations.
\newblock {\em IEEE Transactions on pattern analysis and machine intelligence}
  {\bf 11,} 567--585.

\bibitem[\protect\citeauthoryear{Carnell and Carnell}{Carnell and
  Carnell}{2016}]{carnell2016package}
Carnell, R. and Carnell, M.~R. (2016).
\newblock Package ‘lhs’.
\newblock {\em CRAN. https://cran. rproject. org/web/packages/lhs/lhs. pdf} .

\bibitem[\protect\citeauthoryear{Food, Administration, et~al\mbox{.}}{Food
  et~al.}{2020}]{food2020interacting}
Food, Administration, D., et~al. (2020).
\newblock Interacting with the fda on complex innovative trial designs for
  drugs and biological products.
\newblock {\em FDA} .

\bibitem[\protect\citeauthoryear{Goodfellow, Bengio, and Courville}{Goodfellow
  et~al.}{2016}]{goodfellow2016deep}
Goodfellow, I., Bengio, Y., and Courville, A. (2016).
\newblock {\em Deep learning}.
\newblock MIT press.

\bibitem[\protect\citeauthoryear{Green, Liu, and O’Sullivan}{Green
  et~al.}{2002}]{green2002factorial}
Green, S., Liu, P.-Y., and O’Sullivan, J. (2002).
\newblock Factorial design considerations.
\newblock {\em Journal of Clinical Oncology} {\bf 20,} 3424--3430.

\bibitem[\protect\citeauthoryear{Han, Ren, Wick, Abrey, Das, Jin, and
  Reardon}{Han et~al.}{2014}]{han2014progression}
Han, K., Ren, M., Wick, W., Abrey, L., Das, A., Jin, J., and Reardon, D.~A.
  (2014).
\newblock Progression-free survival as a surrogate endpoint for overall
  survival in glioblastoma: a literature-based meta-analysis from 91 trials.
\newblock {\em Neuro-oncology} {\bf 16,} 696--706.

\bibitem[\protect\citeauthoryear{Hobbs, Barata, Kanjanapan, Paller, Perlmutter,
  Pond, Prowell, Rubin, Seymour, Wages, et~al\mbox{.}}{Hobbs
  et~al.}{2019}]{hobbs2019seamless}
Hobbs, B.~P., Barata, P.~C., Kanjanapan, Y., Paller, C.~J., Perlmutter, J.,
  Pond, G.~R., Prowell, T.~M., Rubin, E.~H., Seymour, L.~K., Wages, N.~A.,
  et~al. (2019).
\newblock Seamless designs: current practice and considerations for early-phase
  drug development in oncology.
\newblock {\em JNCI: Journal of the National Cancer Institute} {\bf 111,}
  118--128.

\bibitem[\protect\citeauthoryear{Hornik}{Hornik}{1991}]{hornik1991approximation}
Hornik, K. (1991).
\newblock Approximation capabilities of multilayer feedforward networks.
\newblock {\em Neural networks} {\bf 4,} 251--257.

\bibitem[\protect\citeauthoryear{Husmann, Lange, and Spiegel}{Husmann
  et~al.}{2017}]{husmann2017r}
Husmann, K., Lange, A., and Spiegel, E. (2017).
\newblock The r package optimization: Flexible global optimization with
  simulated-annealing.

\bibitem[\protect\citeauthoryear{Iasonos, G{\"o}nen, and Bosl}{Iasonos
  et~al.}{2015}]{iasonos2015scientific}
Iasonos, A., G{\"o}nen, M., and Bosl, G.~J. (2015).
\newblock Scientific review of phase i protocols with novel dose-escalation
  designs: how much information is needed?
\newblock {\em Journal of Clinical Oncology} {\bf 33,} 2221.

\bibitem[\protect\citeauthoryear{Jenkins, Stone, and Jennison}{Jenkins
  et~al.}{2011}]{jenkins2011adaptive}
Jenkins, M., Stone, A., and Jennison, C. (2011).
\newblock An adaptive seamless phase ii/iii design for oncology trials with
  subpopulation selection using correlated survival endpoints.
\newblock {\em Pharmaceutical statistics} {\bf 10,} 347--356.

\bibitem[\protect\citeauthoryear{Jones, Attia, Mehta, Liu, Sankhala, Robinson,
  Ravi, Penel, Stacchiotti, Tap, et~al\mbox{.}}{Jones
  et~al.}{2017}]{jones2017tappas}
Jones, R.~L., Attia, S., Mehta, C.~R., Liu, L., Sankhala, K.~K., Robinson,
  S.~I., Ravi, V., Penel, N., Stacchiotti, S., Tap, W.~D., et~al. (2017).
\newblock Tappas: An adaptive enrichment phase 3 trial of trc105 and pazopanib
  versus pazopanib alone in patients with advanced angiosarcoma (aas).
\newblock {\em J. Clin. Oncol.} {\bf 35,} TPS11081.

\bibitem[\protect\citeauthoryear{Kirkpatrick, Gelatt, and Vecchi}{Kirkpatrick
  et~al.}{1983}]{kirkpatrick1983optimization}
Kirkpatrick, S., Gelatt, C.~D., and Vecchi, M.~P. (1983).
\newblock Optimization by simulated annealing.
\newblock {\em Science} {\bf 220,} 671--680.

\bibitem[\protect\citeauthoryear{Leshno, Lin, Pinkus, and Schocken}{Leshno
  et~al.}{1993}]{leshno1993multilayer}
Leshno, M., Lin, V.~Y., Pinkus, A., and Schocken, S. (1993).
\newblock Multilayer feedforward networks with a nonpolynomial activation
  function can approximate any function.
\newblock {\em Neural networks} {\bf 6,} 861--867.

\bibitem[\protect\citeauthoryear{McKay, Beckman, and Conover}{McKay
  et~al.}{2000}]{mckay2000comparison}
McKay, M.~D., Beckman, R.~J., and Conover, W.~J. (2000).
\newblock A comparison of three methods for selecting values of input variables
  in the analysis of output from a computer code.
\newblock {\em Technometrics} {\bf 42,} 55--61.

\bibitem[\protect\citeauthoryear{Mehta, Liu, and Theuer}{Mehta
  et~al.}{2019}]{mehta2019adaptive}
Mehta, C., Liu, L., and Theuer, C. (2019).
\newblock An adaptive population enrichment phase iii trial of trc105 and
  pazopanib versus pazopanib alone in patients with advanced angiosarcoma
  (tappas trial).
\newblock {\em Annals of Oncology} {\bf 30,} 103--108.

\bibitem[\protect\citeauthoryear{Michael and Schucany}{Michael and
  Schucany}{2002}]{michael2002}
Michael, J. and Schucany, W. (2002).
\newblock The mixture approach for simulating new families of bivariate
  distributions with specified correlations.
\newblock {\em The American Statistician} {\bf 56,} 48--54.

\bibitem[\protect\citeauthoryear{Niewczas, Kunz, and K{\"o}nig}{Niewczas
  et~al.}{2019}]{niewczas2019interim}
Niewczas, J., Kunz, C.~U., and K{\"o}nig, F. (2019).
\newblock Interim analysis incorporating short-and long-term binary endpoints.
\newblock {\em Biometrical Journal} {\bf 61,} 665--687.

\bibitem[\protect\citeauthoryear{Pallmann, Bedding, Choodari-Oskooei, Dimairo,
  Flight, Hampson, Holmes, Mander, Odondi, Sydes, et~al\mbox{.}}{Pallmann
  et~al.}{2018}]{pallmann2018adaptive}
Pallmann, P., Bedding, A.~W., Choodari-Oskooei, B., Dimairo, M., Flight, L.,
  Hampson, L.~V., Holmes, J., Mander, A.~P., Odondi, L., Sydes, M.~R., et~al.
  (2018).
\newblock Adaptive designs in clinical trials: why use them, and how to run and
  report them.
\newblock {\em BMC medicine} {\bf 16,} 1--15.

\bibitem[\protect\citeauthoryear{Rasmussen}{Rasmussen}{2003}]{rasmussen2003gaussian}
Rasmussen, C.~E. (2003).
\newblock Gaussian processes in machine learning.
\newblock In {\em Summer school on machine learning}, pages 63--71. Springer.

\bibitem[\protect\citeauthoryear{Razavi, Jakeman, Saltelli, Prieur, Iooss,
  Borgonovo, Plischke, Piano, Iwanaga, Becker, et~al\mbox{.}}{Razavi
  et~al.}{2021}]{razavi2021future}
Razavi, S., Jakeman, A., Saltelli, A., Prieur, C., Iooss, B., Borgonovo, E.,
  Plischke, E., Piano, S.~L., Iwanaga, T., Becker, W., et~al. (2021).
\newblock The future of sensitivity analysis: an essential discipline for
  systems modeling and policy support.
\newblock {\em Environmental Modelling \& Software} {\bf 137,} 104954.

\bibitem[\protect\citeauthoryear{Spall}{Spall}{2005}]{spall2005introduction}
Spall, J.~C. (2005).
\newblock {\em Introduction to stochastic search and optimization: estimation,
  simulation, and control}.
\newblock John Wiley \& Sons.

\bibitem[\protect\citeauthoryear{Thorlund, Haggstrom, Park, and Mills}{Thorlund
  et~al.}{2018}]{thorlund2018key}
Thorlund, K., Haggstrom, J., Park, J.~J., and Mills, E.~J. (2018).
\newblock Key design considerations for adaptive clinical trials: a primer for
  clinicians.
\newblock {\em BMJ} {\bf 360,} k698.

\bibitem[\protect\citeauthoryear{Vanderbeek, Ventz, Rahman, Fell, Cloughesy,
  Wen, Trippa, and Alexander}{Vanderbeek
  et~al.}{2019}]{vanderbeek2019randomize}
Vanderbeek, A.~M., Ventz, S., Rahman, R., Fell, G., Cloughesy, T.~F., Wen,
  P.~Y., Trippa, L., and Alexander, B.~M. (2019).
\newblock To randomize, or not to randomize, that is the question: using data
  from prior clinical trials to guide future designs.
\newblock {\em Neuro-oncology} {\bf 21,} 1239--1249.

\end{thebibliography}
\end{document}